\documentclass[sigconf, nonacm, screen]{acmart}
\usepackage{listings}
\usepackage{balance}
\usepackage[justification=centerlast, skip=12pt]{caption}
\usepackage{amsmath}
\usepackage{tabularx}
\makeatletter
\newif\if@restonecol
\makeatother

\usepackage{algorithm,algpseudocode}
\usepackage{underscore}
\newcommand{\minisec}[1]{\noindent\textbf{#1.}}

\newcommand{\reactdb}{\textsc{ReactDB}}

\setcopyright{rightsretained}
\acmDOI{}
\acmISBN{XXX-X-XXXXX-XXX-X}
\acmConference[]{}{}{}
\acmYear{}
\begin{document}
\title[Evaluation of Intra-Transaction Parallelism in Actor-RDBS]{An Evaluation of Intra-Transaction Parallelism in Actor-Relational Database Systems}

\author{Vivek Shah}
\affiliation{
\institution{Deon Digital Denmark A/S, Denmark}
}
\email{bonii.vivek@gmail.com}
\authornote{Work done while the author was affiliated with the University of Copenhagen}

\author{Marcos Antonio Vaz Salles}
\affiliation{
\institution{University of Copenhagen, Denmark}
}
\email{vmarcos@di.ku.dk}

\begin{abstract}
Over the past decade, we have witnessed a dramatic evolution in main-memory capacity and multi-core parallelism of server hardware. To leverage this hardware potential, multi-core in-memory OLTP database systems have been extensively re-designed. The core objective of this re-design has been scaling up sequential execution of OLTP transactions, wherein alternative database architectures have been explored to eliminate system bottlenecks and overheads impeding inter-transaction parallelism to fully manifest. However, intra-transaction parallelism has been largely ignored by this previous work. We conjecture this situation to have developed because OLTP workloads are sometimes deemed to have far too little intra-transactional parallelism and, even when this kind of parallelism is available, program analyses to recognize it in arbitrary stored procedures are considered too brittle to be used as a general tool. Recently, however, a new concept of actor-relational database systems has been proposed, raising hopes that application developers can specify intra-transaction parallelism by modeling database applications as communicating actors. In this scenario, a natural question is whether an actor-relational database system with an asynchronous programming model can adequately expose the intra-transaction parallelism available in application logic to modern multi-core hardware. Towards that aim, we conduct in this paper an experimental evaluation of the factors that affect intra-transaction parallelism in the prototype actor-relational database system \reactdb, with an OLTP application designed with task-level parallelism in mind, and with the system running on a multi-core machine.
\end{abstract}

\maketitle

\section{Introduction}
Over the past decade there has been a revolution in the hardware landscape and consequently in database system design. As the gains of single-threaded performance of processors dwindled, chip designers have responded by adding processing power in the form of more cores in a processor. In this new continuation of Moore's law, machines with growing numbers of processor cores have become increasingly affordable over time. This trend has also been witnessed in the capacity growth and price decline of main-memory, thus making a single-node multi-core machine today more powerful and cheaper than a cluster of machines two decades back. The database community has responded to this challenge by re-architecting database systems for this brave new world of large main memory and multi-core machines~\cite{Stonebraker:2007:EAE:1325851.1325981, Johnson:2009:SSS:1516360.1516365}. The primary objective of this response, particularly in OLTP DBMS, has been improving transactional scaleup or high-volume processing of sequential transactions. 

\begin{sloppypar}
\textbf{We observe that the potential of speeding up OLTP workloads by intra-transaction parallelism in multi-core in-memory DBMS remains an understudied problem.} Speedup of OLTP workloads has not been traditionally considered worthwhile by OLTP DBMS designers, supposedly because they assume either: (1) in the absence of complex application logic in OLTP transactions, insufficient work is available for parallelization, or~(2) in the presence of complex application logic intermixed with write-heavy data accesses, inferring parallelism to benefit performance while preserving serializability is not in general possible. The latter limitation in inferring parallelism arises from abstracting application logic as stored procedures~\cite{rowes87:storedprocedures}, which are used to ship arbitrary programs to the database system. Stored procedures do not provide any construct to specify task-level parallelism that the database system could exploit. 
\end{sloppypar}

A recent proposal that addresses this problem is to enrich the programming abstraction of stored procedures with modular and asynchronous programming constructs towards actor-relational database systems~\cite{0001S18:ReactDB}. 
The aforementioned work argues for the use of a logical, relational actor (reactor) programming model to: (1)~model applications in terms of actors encapsulating relational state for modularity and locality, and (2)~leverage asynchronous procedure call mechanisms to specify available task parallelism in the application logic. The actor-relational database system prototype \reactdb, proposed in this previous work, ensures that the execution of application logic is carried out under transactional guarantees, freeing application developers to write parallel programs without having to reason about complex synchronization and partial failures for correctness. The work thus argues for marrying actor programming features with the state management capabilities of relational database systems to enable application developers to reap the benefits of application-level modularity and asynchrony, while maintaining insulation from concurrency and failure issues.

Although \reactdb\ demonstrated promising results in leveraging asynchrony in transactions, it has been evaluated in the context of classic OLTP benchmarks, such as TPC-C and YCSB, with almost non-existent application logic. Consequently, the evaluation of \reactdb\ confirmed the known superiority of pursuing a sequential execution strategy for individual OLTP transactions instead of attempting to exploit their available intra-transaction parallelism. Nevertheless, \reactdb's evaluation also provided initial evidence that OLTP transactions having more complex application logic exhibiting sufficient task-level parallelism \emph{can} benefit from the asynchronous programming model provided by reactors. Yet, the evaluation presented in~\cite{0001S18:ReactDB} lacked a detailed, systematic analysis of the factors that affect such intra-transaction parallelism. \textbf{In the present paper, we study these factors and their interplay with the implementation mechanisms in \reactdb\ for OLTP applications with sufficient task-parallel application logic.}

Towards this aim, we first identify the classic factors that affect the performance of parallel programs such as those whose effects we study in our evaluation. We then analyze the system implementation details of \reactdb\ that have been designed to leverage intra-transaction parallelism in these programs. In the absence of well-established OLTP benchmarks with sufficiently complex application logic, we base our evaluation study on a newly proposed SmartMart application~\cite{ShahS17:Actordb}, which models a simplified future IoT supermarket application for next-generation self-checkout~\cite{PlanetMoneySelfCheckout, AmazonGo}. The application attempts to model an upcoming class of OLTP applications with higher computational footprint for trend prediction interspersed with data accesses. The application logic is modeled through a mix of read-mostly transactions~\cite{WangJFP17:SSN-ReadMostly}, including statistical calculations for trend prediction intermixed with conditional statements, aggregation operations, and writes to the database. We vary the deployment of the system as well as inputs to the application procedures to observe the effects of the factors that affect the performance of transactional parallel programs.

In our evaluation, we observe that classic factors affecting parallelism clearly manifest in the measured behavior for intra-transaction parallelism in the studied application. The results indicate that an actor-relational database system can implement intra-transactional parallelism with minimal overhead, and in a way that allows application developers to exploit such parallelism in modern multi-core machines. Based on our evaluation, we believe the time is ripe for investigations of new system designs exploiting both intra-transaction parallelism and inter-transaction parallelism in application benchmarks with sufficient task-parallel application logic. As opposed to focusing solely on scalability of sequential OLTP transactions and associated microarchitectural effects, our work illustrates that the combination of modern hardware and new trends in application workloads open up new possibilities for research in in-memory transactional OLTP database systems. 

In short, we make the following contributions in this paper:
\begin{enumerate}
  \item We study factors that affect intra-transaction parallelism and their effects on a benchmark with complex application logic in the \reactdb\ actor-relational database system prototype. This study includes the observation of possible speedups originating from exploiting the intra-transaction parallelism available in application logic. 
  \item We discuss in detail the implementation of internal components that facilitate intra-transactional parallelism in \reactdb\ and illustrate through our experiments that the associated overheads introduced in the benchmark studied are minimal.
\end{enumerate}

The remainder of this paper is organized as follows. In Section~\ref{sec:background}, we briefly present the classic factors that affect parallel program performance, the actor-relational database system \reactdb, and the SmartMart application used in our evaluation. In Section~\ref{sec:impl}, we drill down into the design and implementation features of \reactdb\ that ensure that the benefits of intra-transaction parallelism in application programs are maintained by the system. In Section~\ref{sec:eval}, we evaluate the system by varying the application parameters and \reactdb\ deployment configurations to quantify the effects of the parallel programming factors enunciated earlier, followed by a holistic discussion of the observed results and their implications. Finally, we discuss related evaluation work in Section~\ref{sec:related} and conclude.

\section{Background}
\label{sec:background}
In this section, we first review the classic factors that affect performance of parallel programs, whose effects we evaluate in the context of an actor-relational database system (Section~\ref{sec:background:factors}). We then provide background on actor-relational database systems and the system chosen for our experimental evaluation, namely \reactdb~\cite{0001S18:ReactDB} (Section~\ref{sec:reactdb:overview}). Finally, we briefly describe the application employed in this paper, SmartMart~\cite{ShahS17:Actordb} (Section~\ref{sec:smartmart:overview}). 

\subsection{Factors Affecting Parallelism}
\label{sec:background:factors}
There are several classic factors that affect the performance of parallel programs~\cite{HennessyPatterson19}, the main of which we recap below.
\subsubsection{Overhead}
Any parallel program is affected by the cost paid by its execution overheads. Examples of overhead comprise: (1)~\emph{Startup costs} that a program has to pay to initiate the parallel computations, including the cost of communication with the parallel processing units necessary to dispatch the work to be performed; (2)~\emph{Transactional overheads} that a program has to pay while running in a database context to obtain transactional guarantees. Some of the overheads can be potentially parallelized, e.g., commit protocol, transaction management overheads, communication cost to send results back from the parallel processing units, while others cannot, e.g., communication cost to initiate parallel computations.  
\subsubsection{Parallelizable Work and Dependencies}
Any parallel program is governed by Amdahl's law~\cite{Amdahl67:Amdahls-law}, which prescribes that the performance of a parallel program is governed by its amount of parallelizable work, i.e., the fraction of total work that would benefit from parallel execution. For a program having a small amount of parallelizable work, the gains from parallelism quickly diminish with increase in parallel resources. As a result, the nature of the program and its functional decomposition, including dependencies among tasks, has a direct influence on the parallelism benefits that can be achieved. In particular, parallelism will not benefit a sequential program, while an embarrassingly parallel program can expect almost linear performance improvements from corresponding increase in parallel resources. Note that dependencies have a direct influence on the synchronization that must be exercised in the program, leading to communication overheads.
\subsubsection{Interference}
Typically, parallel programs are written without consideration for the amount or utilization level of parallel processing resources. In practice, however, when resources are shared for the processing of parallel programs, slowdowns can occur. Resource sharing can happen due to (1)~execution of multiple parallel programs at the same time, or (2)~lack of available resources to sustain the amount of parallelism specified in the program.
\subsubsection{Skew}
In order to get the maximum gain from parallelism, a parallel program must keep all the parallel resources busy at all times. This requires that all tasks executed on the parallel resources have almost the same processing time. This turns out to be especially important as parallel resources are added, since the execution time of the parallel program can become dominated by that of its slowest task. As a result, any imbalance, i.e., skew, in the execution time of the tasks of a parallel program can negatively affect its gains from parallelism.
\subsection{Actor-Relational Database Systems and \reactdb}
\label{sec:reactdb:overview}
\begin{figure}[!t]
\centering
    \scalebox{0.3}[0.29]{\includegraphics{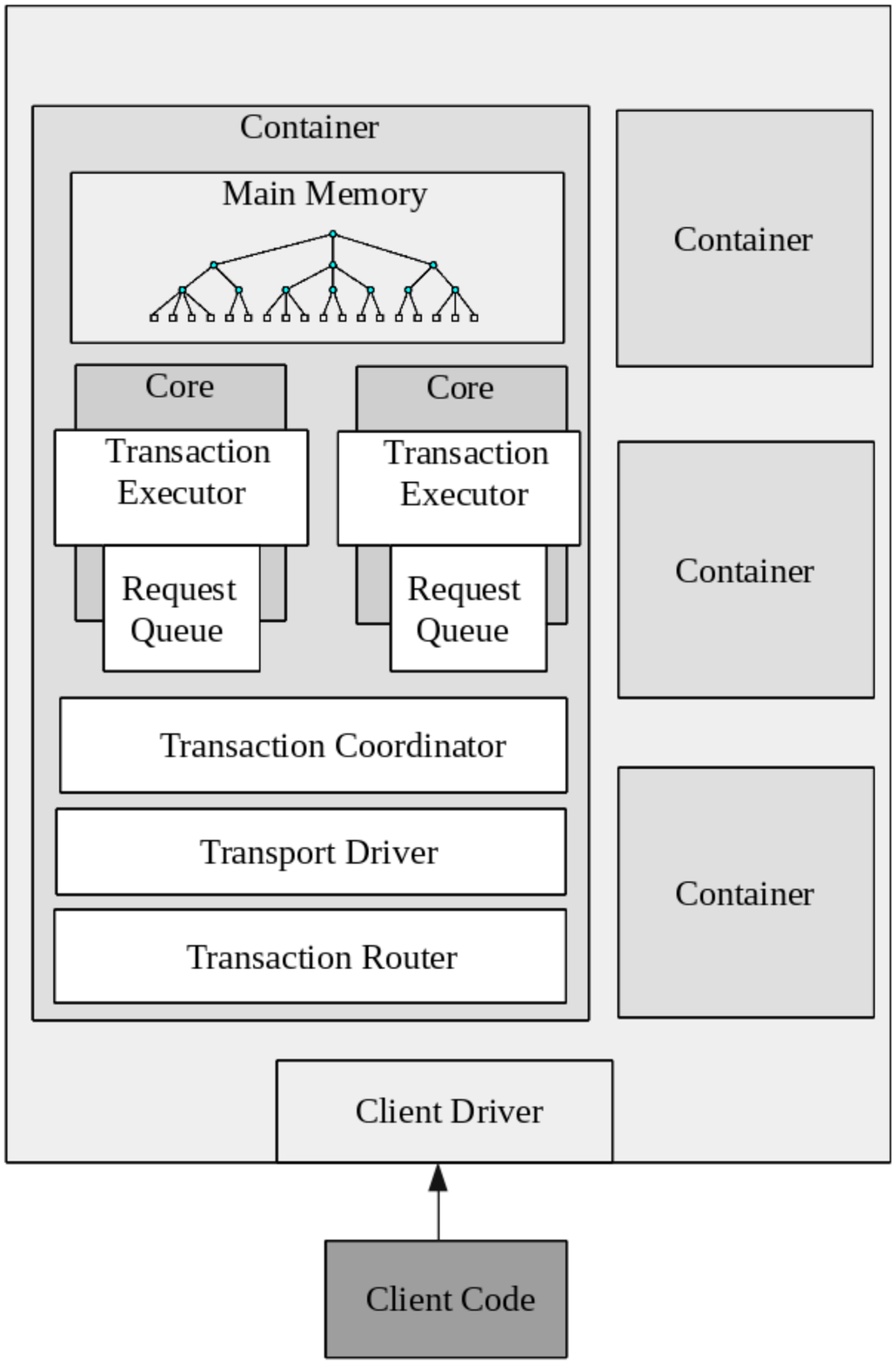}} 
   \vspace{-2ex}
  \caption{\reactdb's architecture~\cite{0001S18:ReactDB}.}
   \vspace{-2ex}
  \label{arch:reactdb}
\end{figure}

Over the last decade, in-memory database systems have received widespread research interest in order to meet the stringent performance demands of online data-intensive applications by utilizing the potential of modern hardware. Despite the criticality of the usage of stored procedures for specifying OLTP applications to leverage performance in modern in-memory DBMS, they are widely perceived as a misfit for modern application needs and hence shunned by application developers~\cite{pavlo17:main-memory-db-problems}. At the same time, actor programming models~\cite{0066897:actorModelBookAgha} and runtimes~\cite{akka-web-documentation, scala-web-documentation, erlang-web-documentation} have seen increased adoption to program stateful applications~\cite{stateful-service-architectures}, even though they lack robust state management features. Observing these technological trends, recent work has proposed the integration of actor programming models in in-memory database systems~\cite{0001S18:ReactDB, ShahS17:Actordb}. This work advocates bringing the benefits of modularity and performance from actors to in-memory database systems. Within this context, the possibility of exploiting parallelism through asynchronicity of communication among actors while maintaining transactional guarantees arises. Even though prior work in actor-relational database systems has shown promising performance results with asynchrony in transactions~\cite{0001S18:ReactDB}, the effect of the multiple factors affecting parallelism still needs to be explored in these systems for OLTP applications having sufficient task parallelism.

Towards the latter, we base this study on \reactdb, an in-memory actor-relational database system prototype designed for multi-core machines. The system provides a logical, actor-oriented programming model~\cite{0001S18:ReactDB}. In \reactdb, a logical actor is referred to as a relational actor, or \emph{reactor}, because the state of an actor is abstracted using relations. Reactors are purely an application-defined construct and exist for the lifetime of the application. Communication among reactors is achieved by asynchronous nested function invocations. In more detail, function invocations on reactors are asynchronous calls returning promises~\cite{Liskov:1988:PLS:53990.54016}, referred in the remainder as \emph{futures}, and can cause nested function invocations on other reactors. However, any function invocation on a reactor is guaranteed to be atomic and serializable. Furthermore, every function invocation is directed to a particular reactor by specifying a unique, application-defined \emph{reactor name}. \reactdb\ does not at present provide durability guarantees. Moreover, application programs in \reactdb\ are written using the reactor programming model directly in C++, where the state of reactors is abstracted through indices supporting interactions against a record-manager interface.
Despite these limitations, the feature set offered by \reactdb\ is sufficient for evaluating the potential benefits of intra-transaction parallelism achieved through asynchronicity in an actor-relational database system, which is the focus of this paper. 

\reactdb\ has been designed to allow for flexible specification of underlying database architecture at deployment time across the extremes of shared-nothing and shared-everything without necessitating any changes to application code. To achieve this flexibility, \reactdb\ virtualizes database architecture by abstracting the notion of sharing of memory and computational resources using \emph{database containers} (see Figure \ref{arch:reactdb}). Computational resources in a database container can access its shared-memory region under a concurrency control protocol. Currently, \reactdb\ utilizes the optimistic concurrency control (OCC) implementation of Silo~\cite{Tu:2013:STM:2517349.2522713} for guaranteeing serializable single-container transaction execution. In addition to the OCC protocol, a multi-container transaction employs a two-phase commit (2PC) protocol between the containers spanned by the transaction. To abstract computational resources, \reactdb\ assigns \emph{transaction executors} to containers. A transaction executor is implemented by a thread pool, pinned to a core, employing a cooperative scheduling strategy. To create a deployment, \reactdb\ requires the specification of the mapping of transaction executors to containers and reactors to transaction executors, such that a reactor can be mapped to only one container. In case a reactor is mapped to more than one transaction executor in a given container, a \emph{transaction router} selects the target destination executor for a function call based on a user-configured policy, e.g., affinity-based or round-robin. We utilize this flexible deployment feature of containers in our experimental evaluation in Section~\ref{sec:eval} to deploy \reactdb\ so as to properly take advantage of the parallelism of the application benchmark described in the following.

\subsection{SmartMart Application Description}
\label{sec:smartmart:overview}
\begin{figure}
\centering
  \includegraphics[width=0.99\linewidth]{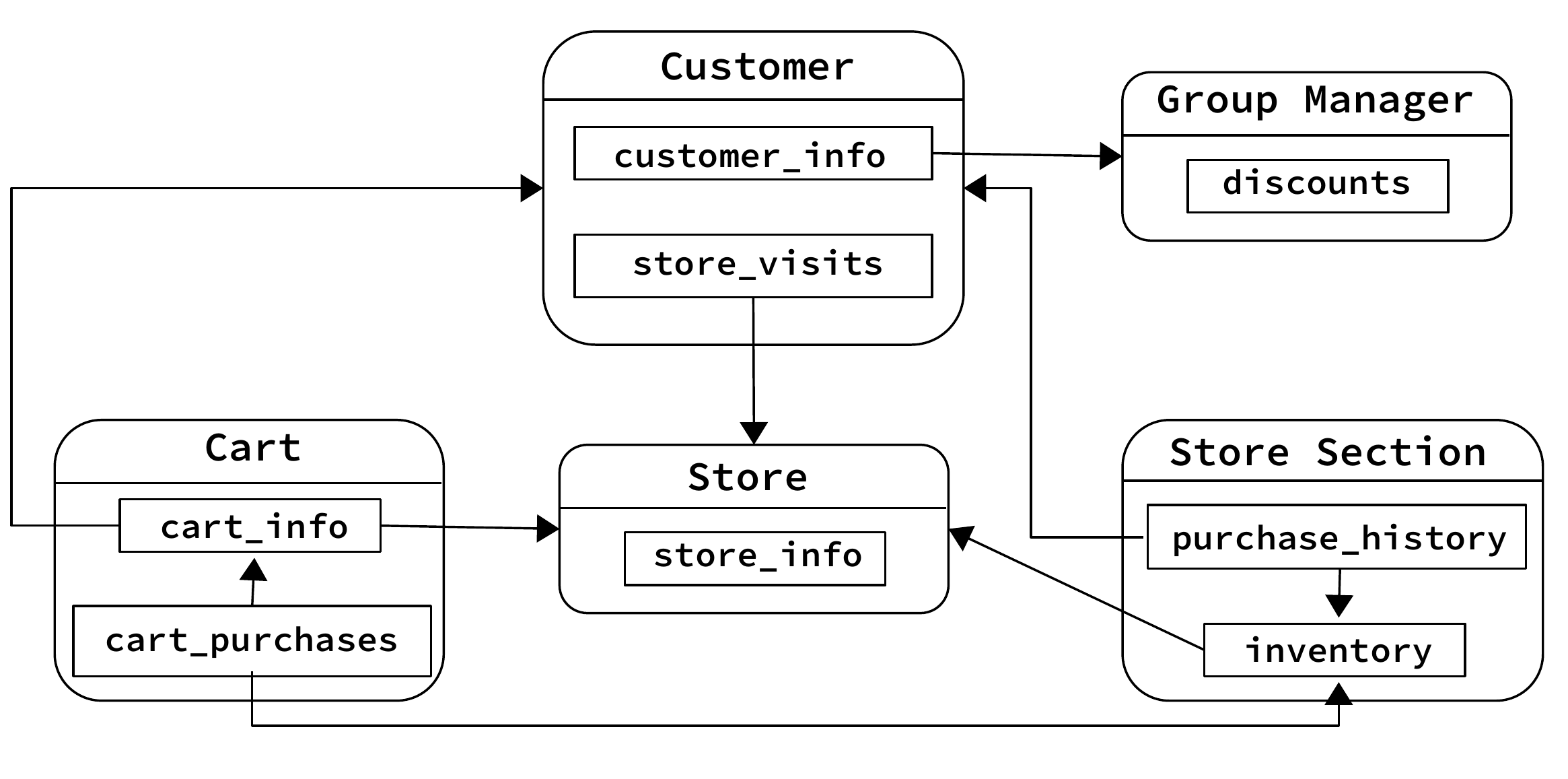}
  \vspace{-1ex}
  \caption{SmartMart schema comprising of actor types, encapsulated relations, and their dependencies}
   \vspace{-1ex}
  \label{fig:smartmart:schema}
\end{figure}

\begin{table*}
   \begin{centering}
    \begin{tabularx}{\linewidth}{|c|X|X|}
      \hline
      Transaction Class & Description & Transaction Footprint \\ \hline
      \texttt{add\_items} & Register addition of items in a cart for a customer. Consists of nested function calls spanning \texttt{Cart, Customer, Group\_Manager, and Store\_Section} reactors to look up a customer's marketing group and then the prices of added items and their fixed discounts based on the customer's marketing group. & Read/write transaction. Has low asynchronicity across the nested function calls to reactors due to control-flow dependencies and the limited amount of work in the nested function calls.\\ \hline
      \texttt{checkout} & Buy the contents of a cart for a customer and terminate the session. Consists of nested function calls spanning \texttt{Cart}, \texttt{Store\_Section} and \texttt{Customer} reactors to update the inventory of the purchased cart items and compute the variable discount for each item using trend prediction over a window of its latest purchases across the store sections. Prices and discounts are then aggregated for the total purchase amount, and finally the customer's visit is recorded. & Read-mostly~\cite{WangJFP17:SSN-ReadMostly} transaction, with computational footprint strongly affected by the trend prediction algorithm employed. Has high asynchronicity across the nested function calls to \texttt{Store\_Section} reactors due to lack of control-flow dependencies and presence of substantial work for variable discount computation. \\ \hline
    \end{tabularx}
    \vspace{1ex}
    \caption{SmartMart Transaction Classes}
    \vspace{-1ex}
    \label{tab:smartmart:transactions}
   \end{centering}
\end{table*}

The SmartMart application~\cite{ShahS17:Actordb} was designed to model a simplified future IoT supermarket application for next-generation self-checkout~\cite{PlanetMoneySelfCheckout, AmazonGo}. The application models the workflow of a customer inside the supermarket carrying a smart shopping cart that is equipped with sensors to itemize its physical contents and trigger checkout operations. The application defines two different types of discounts on each item in the inventory: (1)~fixed discount and (2)~variable discount. The fixed discount is customized based on the marketing group of the customer, while the variable discount is computed using trend prediction algorithms with the recent history of demand for an item over a pre-defined window as input~\cite{ShahS17:Actordb}. The computational footprint of the trend prediction algorithm makes it particularly attractive for parallelization.

In the application specification~\cite{ShahS17:Actordb}, the functional decomposition of the application across reactors has already been done along with the schema specification for each reactor type and the declarative queries to interact with the encapsulated reactor state. We use the same decomposition of reactors in this paper. The application consists of five reactor types and eight relations as outlined in the schema definition in Figure \ref{fig:smartmart:schema}. Note that the schema consists of reactor types and their encapsulated relations. Not only are there 1:N dependencies between the relations, but also between relations and reactor types (with reference to reactor names) . The workload comprises of a workflow of (1)~adding items and (2)~checkout, simulating the actions of customers in the supermarket as mentioned earlier and described in detail in Table \ref{tab:smartmart:transactions}. The throughput and latency metrics are reported for the entire workflow. We ported the declarative queries of the application into handcrafted physical plans over the database index interfaces of \reactdb.

We employ the SmartMart application because it has been specifically designed for actor-relational database systems with task-level parallelism in mind, and thus has the necessary parameters that allows us to vary and observe the effects of the factors outlined in Section~\ref{sec:background:factors}.

\section{Transactional Parallelism Implementation}
\label{sec:impl}
In this section, we discuss the implementation details of \reactdb\ related to transactional parallelism. \reactdb's core abstraction for intra-transaction parallelism is that of asynchronicity in nested function calls, which are provided while maintaining transactional guarantees of atomicity and serializability. We begin by first discussing the thread pool management features in a transaction executor, which are the active computational entities in \reactdb\ (Section~\ref{sec:thread:management}). We then explain the workflow followed by a transaction executor (Section~\ref{sec:txn:executor}), and continue by discussing the implementation of function calls on reactors and the associated dispatch mechanism across containers and transaction executors (Section~\ref{sec:function:call}). Finally, we outline the commit protocol employed by the system (Section~\ref{sec:commit}), before discussing issues of code generation and memory management that help in achieving high performance in \reactdb\ (Section~\ref{sec:others}).

\subsection{Thread Management In Transaction Executors}
\label{sec:thread:management}

\begin{figure}[!t]
  \centering
  \includegraphics[width=0.95\linewidth]{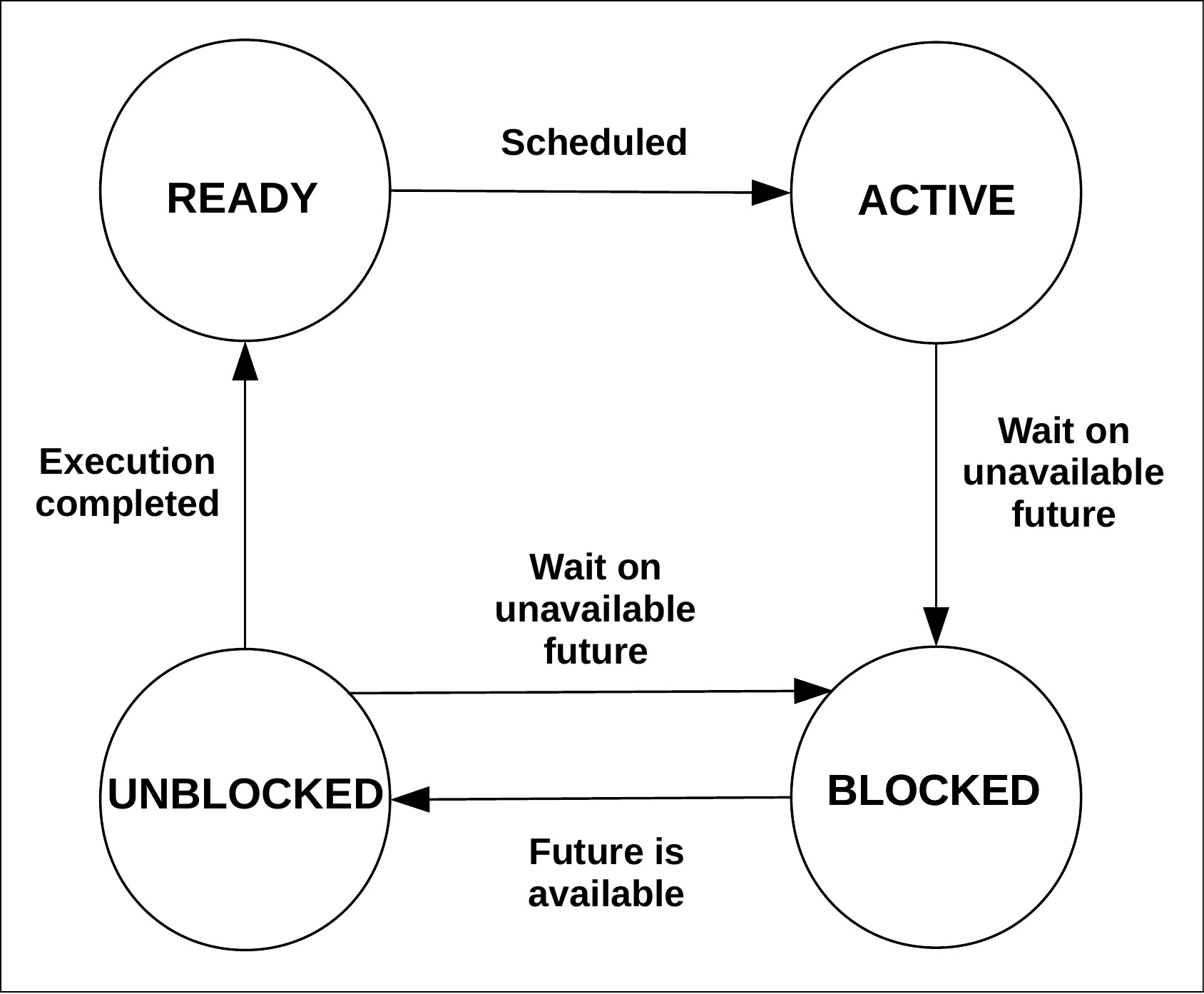}
  \caption{Transaction Executor Thread States and Transitions}
  \label{fig:impl:thread:lifecycle}
\end{figure}

A transaction executor is a thread pool pinned to an actual physical core. Transaction executors are created when \reactdb\ is bootstrapped based on an assignment to containers specified in configuration files. \reactdb\ runs as a single process, and all the threads in a transaction executor after creation wait to be scheduled in order to run. Every transaction executor has its own thread scheduler, which is configured with the number of active threads allowed to run at any point of time. Figure~\ref{fig:impl:thread:lifecycle} depicts the thread states and the events that cause transitions between these states. Every thread in a transaction executor is in one of these four states at any point of time.

After creation, all the threads are in the READY state and wait for permission from the scheduler to run. A scheduler allows a thread to run only if the number of current active threads does not exceed the limit of active threads configured per scheduler. Once a thread gets permission to run, it enters the ACTIVE state and keeps dequeuing and executing (sub-)transaction requests from the transaction executor queue. During the course of execution of (sub-)transactions, the thread can get blocked if it needs to wait for the availability of a future result of a nested sub-transaction. In such a case, the thread transitions to the BLOCKED state and notifies the scheduler. The scheduler then permits another thread in READY state to begin execution, if available. The BLOCKED thread transitions to the UNBLOCKED state once the sub-transaction result it is waiting for becomes available. Until the end of the execution of the current (sub-)transaction, the thread never returns to the ACTIVE state, but can transition back and forth between BLOCKED and UNBLOCKED. When the entire execution of the (sub-)transaction is complete, the thread returns back to the thread pool, transitioning to the READY state. As such, the thread can now once again wait for permission from the scheduler to run. 

These mechanics of thread scheduling aim to ensure that the queue of a transaction executor is continuously drained by active threads up to the limit on thread resources. This minimizes delay in scheduling of transactions and sub-transactions while controlling admission to compute resources allocated to the system. Note that a thread in the ACTIVE state that never needs to wait for an unavailable sub-transaction result will drain the transaction executor queue continuously and not return back to the thread pool. 

\subsection{Transaction Executor Workflow}
\label{sec:txn:executor}
\begin{algorithm}[!t]
  \caption{Transaction Executor Workflow}
  \label{alg:txn-executor-workflow}
  \begin{algorithmic}[1]
  \Procedure{TxnExecutor::executeForever }{sched, queue, coord, container}
    \While{$true$} 
      \State $sched.await()$
      \While{$runningUnblocked$} 
        \State $txn \gets queue.deque()$
        \If{$txn.isRootTxn()$} 
          \State $tid \gets generateTid()$
          \State $txn.setTid(tid)$
          \State $dbCtx \gets container.createDbCtx(tid)$
        \Else
        \State $dbCtx \gets container.getOrCreateDbCtx($
        \State $\;\;\;\;\;\;\;\;\;\;\;\;\;\;\;\;\;\;\;\;\;\;\;\;\;\;\;\;\;\;\;\;\;\;\;\;\;\;\;\;\;\;\;\;\;\;\;\;\;txn.getTid())$
        \EndIf
        \State $txn.setDbCtx(dbCtx)$
        \State $txn.run()$
        \State $success \gets txn.waitForSubTxns()$
        \If{$txn.isRootTxn()$}
          \State $coord.commitOrAbort(txn, container, success)$
        \Else
          \State $txn.dispatchResult()$;
        \EndIf
      \EndWhile
      \EndWhile
      \EndProcedure
  \end{algorithmic}
\end{algorithm}
Algorithm~\ref{alg:txn-executor-workflow} represents the workflow of the threads in the transaction executor. The procedure \texttt{EXECUTEFOREVER} receives as arguments the scheduler for the transaction executor, the transaction executor queue, the container to which the transaction executor belongs, and the transaction coordinator for the container. As explained in the previous section, when the threads in a transaction executor are created, they wait for permission from the scheduler to run (line 3). Once permission is granted by the scheduler, the thread dequeues (sub-)transactions from the transaction executor queue (line 5). Every transaction has a database context that stores necessary concurrency control information such as read and write sets for OCC. This database context is stored in the container, since the context needs to be re-used across all nested sub-transaction executions touching the same container. 

Every database context of a transaction is mapped by a transaction identifier that is generated once when a root transaction begins execution. Even though this identifier is carried along in sub-transaction invocations across containers, every transaction has only one database context in any given container. Lines 6-12 show the creation of transaction identifier, the creation of the database context for a transaction, and the lookup of database contexts for sub-transactions. To avoid conflicts among sub-transactions of the same root transaction, the transaction executor also ensures that a database context in a container is accessed by only one thread at any point of time and conservatively aborts any transactions violating this requirement. The database context is utilized during the execution of (sub-)transaction code (line 15), where the \texttt{run} method abstracts the application logic.

Once the application logic completes, an explicit synchronization is done to wait for the completion of nested children sub-transactions. This guarantees that the parent can only return once all child sub-transactions are complete, recursively (line 16). As such, when all children of a root transaction complete, the transaction execution is complete and a commit protocol must be initiated based on the results of the transaction execution. The commit protocol runs in the transaction coordinator, and ensures that a transaction is aborted if the execution encountered errors and committed otherwise (line 18). Finally, if the transaction is not a root transaction, the result of the sub-transaction must be dispatched back to the caller (line 20). A transaction executor also performs other functionality for memory management and garbage collection that we omit from the pseudocode for simplicity. 

The workflow followed by a transaction executor ensures that transactional contexts are managed properly across multiple containers in support of atomicity and serializability. Note that the workflow does not make any assumptions on patterns of nested asynchronous function calls in application logic and will accept programs with arbitrary nested invocations of functions and synchronization structures, allowing for expression of complex intra-transaction parallelism in application control flow.

\subsection{Asynchronous Function Call Implementation}
\label{sec:function:call}

\begin{algorithm}[!t]
  \caption{Asynchronous Transaction Execution}
  \label{alg:impl:async:functions}
  \begin{algorithmic}[1]
    \Function{Txn::exec }{parentTxn, txn, srcContainer, reactor, args}
      \State $txn.setInput(args)$
      \State $txn.setReactor(reactor)$
      \State $result \gets txn.createResultFuture()$
      \If{$\neg txn.isRootTxn()$}
        \State $parentTxn.addToSubTxns(txn)$
        \State $txn.setTid(parentTxn.getTid())$
      \EndIf
      \State $dstCont \gets srcContainer.GetContainer(reactor)$
      \If{$dstCont = srcContainer$}
          \State $txn.setDbCtx(parentTxn.getDbCtx())$ 
        \State $txn.run()$
      \Else
        \State $txn.setContainer(dstCont)$
        \State $dstContainer.schedule(txn)$
      \EndIf
      \State \Return $result$  
    \EndFunction
   \end{algorithmic}
\end{algorithm}
Having outlined the workflow of a transaction executor in the previous section, we now explain how asynchronous function calls are implemented in \reactdb. Specifically, we outline the steps triggered when the application logic invokes a function call on a particular reactor.\footnote{In the reactor programming model~\cite{0001S18:ReactDB}, such a call corresponds to the use of the construct \texttt{fn(args) on reactor X} in the application logic.}

Algorithm~\ref{alg:impl:async:functions} shows the corresponding pseudocode. The function \texttt{EXEC} captures the dispatch logic of a function call on a reactor, abstracted in \reactdb\ by a (sub-)transaction \texttt{txn}. \texttt{EXEC} receives as arguments a handle to the parent (sub-)transaction from which the (sub-)transaction was invoked (empty for a root transaction), the container where the parent transaction was executing, the reactor on which the (sub-)transaction is invoked, and the arguments to the (sub-)transaction. Lines~2 and~3 store the (sub-)transaction inputs and the (sub-)transaction reactor inside \texttt{txn} so that these values can be later accessed by the transaction logic. Line~4 creates the future result of the transaction that will be later returned at the end of the function (line~17). If \texttt{txn} is a not a root transaction, then its handle is stored in the list of sub-transactions of the parent (lines~5 and~6). This list of sub-transactions is used by the parent to ensure synchronization of its children sub-transaction executions in \texttt{waitForSubTxns} (Algorithm~\ref{alg:txn-executor-workflow}). The identifier of the sub-transaction is also set to that of the parent (sub-)transaction for all non-root transactions (line 7).

\begin{sloppypar}
To execute the sub-transaction, the destination container mapped to the \texttt{reactor} is looked up by consulting the reactor-to-container mapping stored in the source container (line 9). If the destination container is the same as the source container (only relevant for non-root transactions), then the database context of the parent sub-transaction is stored in the child sub-transaction and the method call to the application logic is directly invoked. This action chains the invocation sequence, leading to synchronous execution (lines 10-12). The intuition for this decision is that \reactdb\ wants to eliminate the overhead of rescheduling the sub-transaction when a migration to another container is unnecessary. As a special case, this decision also ensures that a sub-transaction invoked on the same reactor is synchronously executed with minimal overhead.
\end{sloppypar}

If the source container differs from the destination container, the destination container is stored in the (sub-)transaction for later use during the commit protocol. Then, the transport driver of the source container is invoked to move the \texttt{txn} to the destination container (line 15). In the current implementation of \reactdb, the logic utilizes shared-memory access to directly invoke \texttt{schedule} on the destination container, which looks up the destination transaction executor for the reactor with the aid of the transaction router and enqueues the (sub-)transaction into the corresponding transaction executor queue. 

\begin{algorithm}[!t]
  \caption{Commit Protocol}
  \label{alg:impl:commit}
  \begin{algorithmic}[1]
    \Procedure{TxnCoordinator::CommitOrAbort}{rootTxn, srcContainer, commit}
      \If{$commit$}
        \State $success \gets srcContainer.getDbCtx().validate()$
      \Else
        \State $success \gets false$
      \EndIf
      \State $containers \gets rootTxn.getRemoteContainers(srcContainer)$
      \If{$success$}
        \ForAll{$container \in containers$}
          \State $ success \gets container.getDbCtx().validate()$
          \If{$\neg success$}
            \State \textbf{break}
          \EndIf
        \EndFor
      \EndIf
      \If{$success$}
        \State $srcContainer.getDbCtx().write()$
        \ForAll{$container \in containers$}
          \State $ container.getDbCtx().write()$
        \EndFor
      \Else
        \State $srcContainer.getDbCtx().abort()$  
        \ForAll{$container \in containers$}
          \State $ container.getDbCtx().abort()$
        \EndFor
      \EndIf    
      \State $rootTxn.setCommitStatus(success)$
      \State $rootTxn.dispatchResult()$
      \State $rootTxn.end()$
    \EndProcedure
   \end{algorithmic}
\end{algorithm}
\subsection{Commit Protocol}
\label{sec:commit}
After the transaction logic is executed, \reactdb\ ensures that all the results of the sub-transactions of a root transaction are available. Then, the transaction coordinator initiates a linear two-phase commit (2PC) protocol to either commit or abort the root transaction. Figure~\ref{alg:impl:commit} outlines the implementation of the 2PC protocol in \reactdb. The validation phase of the OCC protocol is executed first in the source container, i.e., where the root transaction was initiated and consequently the transaction coordinator was invoked (line 3). After this validation step, all the remote containers that are spanned by the transaction are looked up recursively through the chain of sub-transactions starting from the root transaction (line 7). 

If validation was successful in the source container, lines 8-15 proceed to run the validation phase of the OCC protocol on each of the remote containers. Following this process, either the write phase (lines 17-20) or the abort phase (lines 22-25) is executed, depending on the combined result of validation. At the end, the commit status of the root transaction is set, so that it can be examined by the caller (line 27). At this point, the execution and commitment of the transaction is complete, so the caller is notified of the transaction completion (line 28) and necessary cleanups are performed for the transaction (line 29). We observe that the linear two-phase commit protocol currently employed in \reactdb\ can become a source of overhead that can affect the gains from parallelization and could potentially be improved to further lower its overhead. 

\subsection{Other Implementation Factors}
\label{sec:others}

In this section, we highlight additional implementation decisions taken during the design of \reactdb\ that contribute to low overhead and high performance, thus accentuating the benefits of intra-transaction parallelism.

\subsubsection{Efficient Code Generation}
\reactdb\ has been designed as a framework heavily using C++11 templates to minimize any dynamic dispatches during execution. In particular, C++ templates are used in implementation throughout the entire code line, including index structures, concurrency control mechanisms, transaction coordinators, transaction executors, transaction routers, schedulers and containers. The application code in stored procedures extends \reactdb\ transaction classes and is thus compiled with \reactdb\ instead of being linked separately, providing the compiler with a large body of program code to optimize. In addition, the compiler has the opportunity to specialize the code generation according to a set of static configuration options of \reactdb. 

\subsubsection{Efficient Memory Management}
In order to prevent bottlenecks in memory allocation, \reactdb\ uses a custom memory allocator for each thread in the transaction executor that pre-allocates memory on the heap and only dynamically allocates memory if pre-allocated memory is exhausted. Furthermore, any memory dynamically allocated is not freed immediately, but reserved by the allocator for later use. Thus, this results in the memory pool being resized dynamically. Transaction execution consumes memory from the memory pool while garbage collection reclaims memory back on completion of transactions. \reactdb\ also makes extensive use of the stack instead of the heap where possible in its implementation.

\section{Evaluation}
\label{sec:eval}
In this section, we evaluate the effects of the classical parallelism factors outlined in Section~\ref{sec:background:factors} on intra-transaction parallelism in actor-relational database systems using the SmartMart application. As opposed to previous results on OLTP benchmarks~\cite{0001S18:ReactDB}, our goal is to study the benefits of exploiting parallelism in a workload that is well-matched conceptually to the actor-relational database approach and has task-level parallelism. More concretely, we aim to answer the following questions:
\begin{itemize}
  \item How does the amount of parallelizable work and the overheads of an actor-relational database system affect the overall speedup of transaction programs (Sections~\ref{sec:varying:scan:size} and~\ref{sec:exp:speedup})? 
  \item How does interference from inter-transaction parallelism affect the benefits of intra-transaction parallelism (Section~\ref{sec:exp:scaleup})? 
  \item How does skew or load imbalance in intra-transaction parallelism affect the benefits experienced by transaction programs (Section~\ref{sec:exp:skew})? 
\end{itemize}
\subsection{Experimental Setup}
\subsubsection{Hardware}
We focus on multi-core hardware and use a machine with two sockets, each with one eight-core 2.6 GHz Intel Xeon E5-2650 v2 processor with two physical threads per core, leading to a total of 32 hardware threads. Each physical core has a private 32~KB L1 data and instruction cache and a private 256~KB L2 cache. All the cores on the same socket share a last-level L3 cache of 20~MB. The machine has 128~GB of RAM, with half the memory attached to each of the two sockets, and runs 64-bit RHEL Linux 3.10.0. 
\subsubsection{Workload}
\label{sec:eval:workload}
As discussed in Section~\ref{sec:smartmart:overview}, we employ the SmartMart application in our experiments and use the same reactor modeling for the application as done previously in its specification~\cite{ShahS17:Actordb}. To simulate the workings of one store, we created eight \texttt{Store\_Section} reactors.~For each, we loaded the \texttt{inventory} relation with 10,000 items and the \texttt{purchase\_history} relation with 300 entries per item for a total of 3,000,000 entries, simulating a history of 120 days where 500 customers on average visit the store per day and buy 50 items each. For simplicity of interpretation of the results, we do not model application logic that is sensitive to the skew in the distribution of item purchases, and leave this extension to future work. We fix the number of \mbox{\texttt{Group\_Manager}} reactors to 10 and vary the number of \texttt{Cart} reactors depending on the experiment. The number of \texttt{Customer} reactors is set to 30 times the number of carts. To calculate the variable discount, we used a statistical function for trend prediction based on the mean and variance over the reverse purchase history $S_{i,t}^{k}$ starting at time $t -1$ of size at most $k$ for each item $i$: 
\begin{equation} 
pred_{q}(S_{i,t}^{k}) = \mu (S_{i,t}^{k}) \quad + \quad \sigma (S_{i,t}^{k}) 
\end{equation}
The entire size allocated after loading was $\sim$3~GB. We note that the prediction function used is distributive and its computation can be supported by database aggregate functions. This is not, however, true in general, as more sophisticated prediction schemes could instead be employed for trend prediction~\cite{RasmussenW2006:GPBook}. We chose a statistical function for trend prediction because its lower computational footprint compared to other learning methods would not completely dwarf the commit overhead.

\subsubsection{System Configuration}
As discussed in Section~\ref{sec:reactdb:overview}, we run our experiments in an actor-relational database system prototype named \reactdb. Relations in reactors are represented by index structures~\cite{MaoKM12:Masstree}. The system guarantees globally serializable execution of programs by using a combination of optimistic concurrency control~\cite{Tu:2013:STM:2517349.2522713} and two-phase commit protocols. The system also supports flexible mapping of reactors to hardware resources, resulting in different possibilities for deployment. 

\begin{sloppypar}
For our experiments, we configured the deployment architecture of our system under the following two modes:\\
\textbf{(1) \textsf{sync}} - In this setting, we configured \reactdb\ to execute the logic of each single transaction sequentially similar to modern in-memory OLTP database systems~\cite{Tu:2013:STM:2517349.2522713}. To achieve this effect, we employed a single container with an affinity-based router and with eight transaction executors for the corresponding physical cores on the first socket of the machine. Each \texttt{Cart} reactor and its associated set of \texttt{Customer} reactors were mapped to a disjoint transaction executor, while all the \texttt{Store\_Section} and \texttt{Group\_Manager} reactors were mapped to all the transaction executors. As a result, each invocation of a root transaction on a \texttt{Cart} reactor is processed by a unique transaction executor, which also executes all sub-transactions even if to different reactors.
Each worker thread generating method invocations on a \texttt{Cart} reactor got mapped to the hyper-threaded core of the corresponding cart, simulating client affinity.\\
\textbf{(2) \textsf{async}} - In this setting, we configured \reactdb\ to take advantage of the intra-transaction parallelism in the benchmark, especially in the \texttt{checkout} transaction. To this end, we utilized the other socket on the machine to create eight containers with affinity-based routers each consisting of one transaction executor for each of the physical cores on the second socket. The eight \texttt{Store\_Section} reactors were mapped to a single container each, and consequently only one transaction executor each. We employed another container similar to the \textsf{sync} setting on the first socket of the machine to map \texttt{Cart}, \texttt{Customer}, and \texttt{Group\_Manager} reactors. As a result, only a sub-transaction invocation to a \texttt{Store\_Section} reactor is dispatched to the corresponding remote transaction executor and container for asynchronous execution, while all sub-transaction invocations to other reactor types are executed synchronously since they are mapped to the same container. We refrained from employing a mapping similar to \texttt{Store\_Section} reactors for the \texttt{Group\_Manager} and \texttt{Customer} reactors, because we were limited by the physical parallelism of the machine and we wanted to avoid hyper-threading effects. In addition, the amount of work in the asynchronous method invocation to these reactors is minimal compared to the \texttt{Store\_Section} reactors responsible for trend prediction, which would potentially increase execution cost considering the overhead of dispatch as opposed to shared-memory operations.
\end{sloppypar}
Since we could only use eight \texttt{Cart} reactors in the \textsf{async} setting, we only used one socket in the \textsf{sync} setting in order to make comparisons across the two settings. We tuned the thread pool sizes for the transaction executors to minimize queuing delays and maximize usage of physical cores.

\subsubsection{Methodology}
A worker runs interactions consisting of (1) \texttt{add\_items}, and on its successful commit, (2) \texttt{checkout}. We measure the average latency and throughput of the entire interaction using an epoch-based measurement strategy~\cite{DBLP:journals/pvldb/DifallahPCC13}. Each epoch consists of two~seconds, and we report averages and standard deviations of successful interactions over 20 epochs. Workers choose customer IDs from a uniform distribution. 
Unless mentioned otherwise, the items and store sections in orders are also chosen from a uniform distribution for a configurable number of store sections and items per store section in the order. 

\begin{figure*}[t]
    \begin{minipage}{0.32\linewidth}
        \centerline{\includegraphics[width=0.95\linewidth]{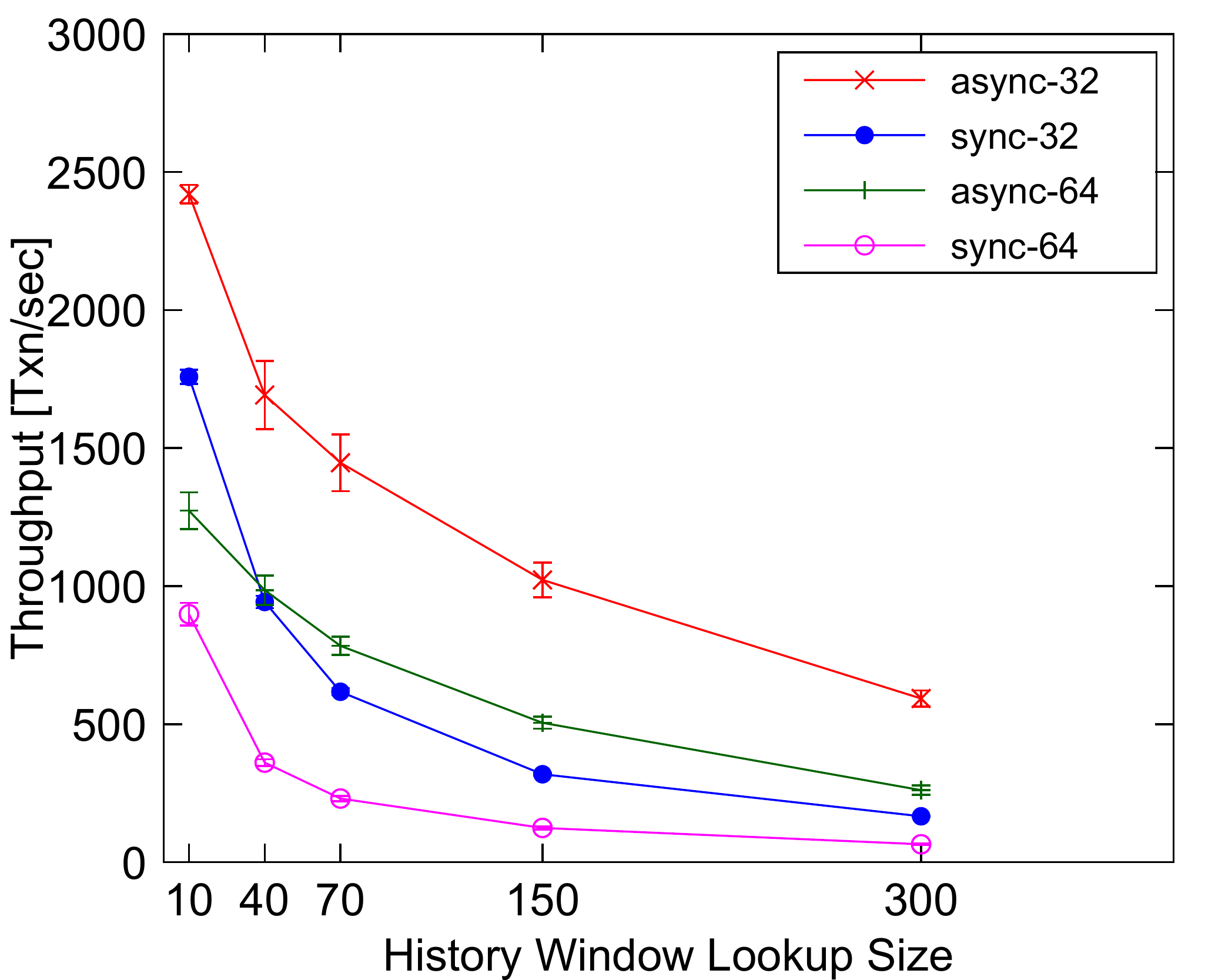}}
        \caption{Effect of varying history window scan size on throughput.} \label{fig:scan:throughput}
    \end{minipage} \hfill
    \begin{minipage}{0.32\linewidth}
        \centerline{\includegraphics[width=0.95\linewidth]{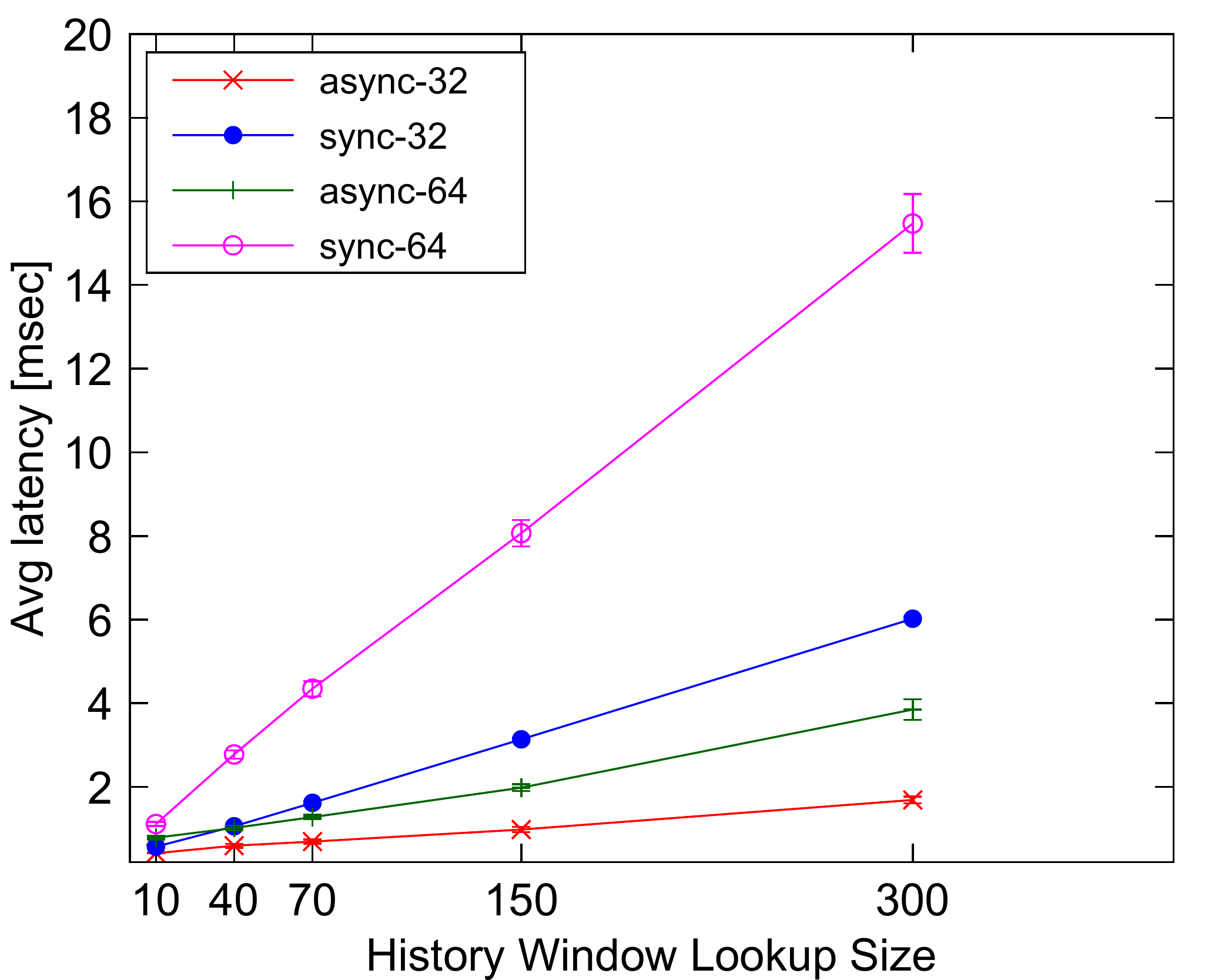}}
        \caption{Effect of varying history window scan size on latency.} \label{fig:scan:latency}
      \end{minipage} \hfill
    \begin{minipage}{0.32\linewidth}
    \centerline{\includegraphics[width=0.95\linewidth]{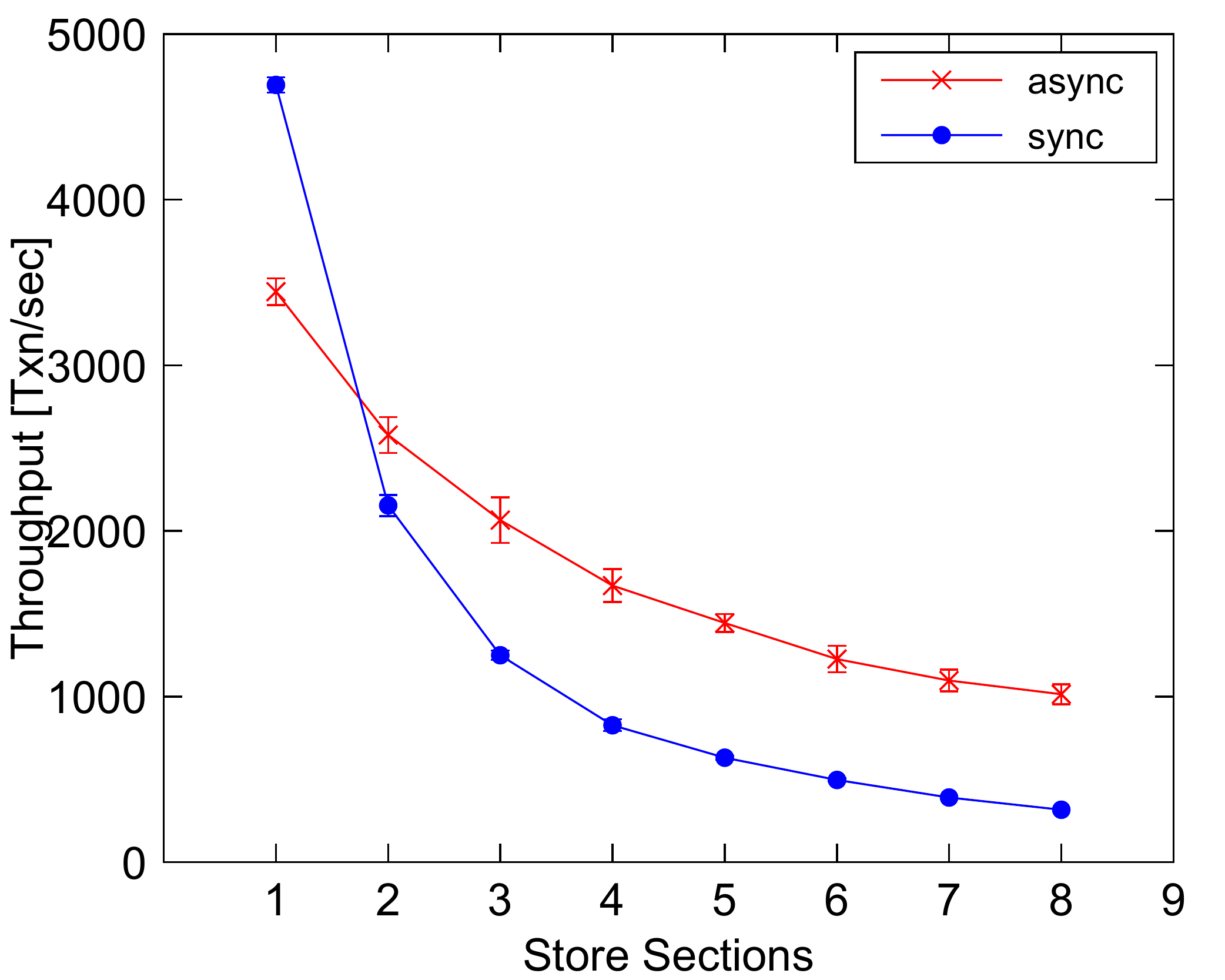}}
    \caption{Effect of varying order size on throughput.} \label{fig:throughput:parallel:scaling}
    \end{minipage} \hfill
\end{figure*}

\subsection{Can asynchrony and parallelism help cope with increasing work in \textsf{SmartMart}?}
\label{sec:varying:scan:size}
In this section, we study the benefits of asynchrony in order to deal with increasing parallelizable work in a transaction. We varied the parallelizable work within a transaction in two ways: (1) by keeping the number of nested asynchronous functions fixed, but varying the amount of work in each of these functions (see \emph{Constant Parallel Degree}); (2) by increasing the number of nested asynchronous functions, but keeping the amount of work in each of them fixed (see \emph{Increasing Parallel Degree}). To avoid interference from inter-transaction parallelism, we run these experiments with a single worker.

\minisec{Constant Parallel Degree}  
In this experiment, we study the effect of varying the amount of parallelizable work in variable discount computation at checkouts on transaction throughput and latency.  The number of store sections remains fixed at eight, and the number of items scanned in the \texttt{purchase\_history} relation for variable discount computation is varied. To avoid skew, the number of items requested from each store section were identical. Finally, to observe the influence of transaction size, we also experiment with two different settings for the number of items requested from each store section, namely either four or eight, thus resulting in a total number of items purchased across all store sections to be either 32 or 64.

Figures~\ref{fig:scan:throughput} and~\ref{fig:scan:latency} depict the throughput and latency results obtained. Since we used a single worker, latency and throughput numbers are almost inverses of each other. We can see that as we increase the history window scan size from 10 to 300, the drop in throughput of \textsf{sync} is higher than that of \textsf{async}. The growing work due to the larger scan sizes and associated mean and standard deviation computations can be parallelized across the store sections in the \textsf{async} version, but not in \textsf{sync}. As a consequence, the curves for the same transaction sizes diverge (this effect can be seen more clearly in Figure~\ref{fig:scan:latency}).

Overall, \texttt{async} dominates \texttt{sync} throughout the experiment, due to its ability to exploit intra-transaction parallelism and additional resources even if at the cost of higher overheads. It is interesting to see that as we vary the history window scan size beyond 40, \textsf{async-64} starts outperforming \textsf{sync-32}, i.e., a sequential transaction is outperformed by an asynchronous transaction of twice the size. \emph{Based on the results of this experiment, we choose a history window scan size of 150 for all future experiments}, because (1) this corresponds to 150 records in the \texttt{purchase\_history} relation, thus calculating a target purchase quantity over 60 days, which is a realistic estimate based on our assumptions of the workload as explained in Section~\ref{sec:eval:workload}, and (2) it allows us to experiment with and demonstrate the effects of intra-transaction parallelism clearly. At this point, the speedup of \texttt{async-64} over \texttt{sync-64} is between four and five, which we discuss further in Section~\ref{sec:exp:speedup}.

\minisec{Increasing Parallel Degree}
In this experiment, we study the effect of increasing both work and asynchronicity in method calls.
We vary the number of store sections from 1 to 8 while keeping the number of items ordered from each section fixed at 4, thus varying the size of the order from 4 to 32. Figure~\ref{fig:throughput:parallel:scaling} shows that the throughput of \textsf{sync} degrades with increasing order size given the sequential execution of the methods. The slope of the curve also decreases with store sections since the increase in the order size is constant, and hence has a smaller impact as the order size grows. By contrast, \textsf{async} has lower throughput when the number of store sections is one, but reaches 3.2x higher throughput than \textsf{sync} for 8 store sections. In the beginning, \textsf{async} suffers from dispatch overheads and lack of sufficient asynchrony as opposed to shared memory accesses in \textsf{sync}. However, as the number of store sections increases, asynchronicity benefits arise since the variable discount computations across store sections during \texttt{checkout} and price lookups during \texttt{add\_items} are overlapped to utilize parallel resources.

\subsection{What speedups can we achieve with increasing parallel resources and parallelizable work in \textsf{SmartMart}?}
\label{sec:exp:speedup}
\begin{figure*}
    \begin{minipage}{0.32\linewidth}
      \centerline{\includegraphics[width=0.95\linewidth]{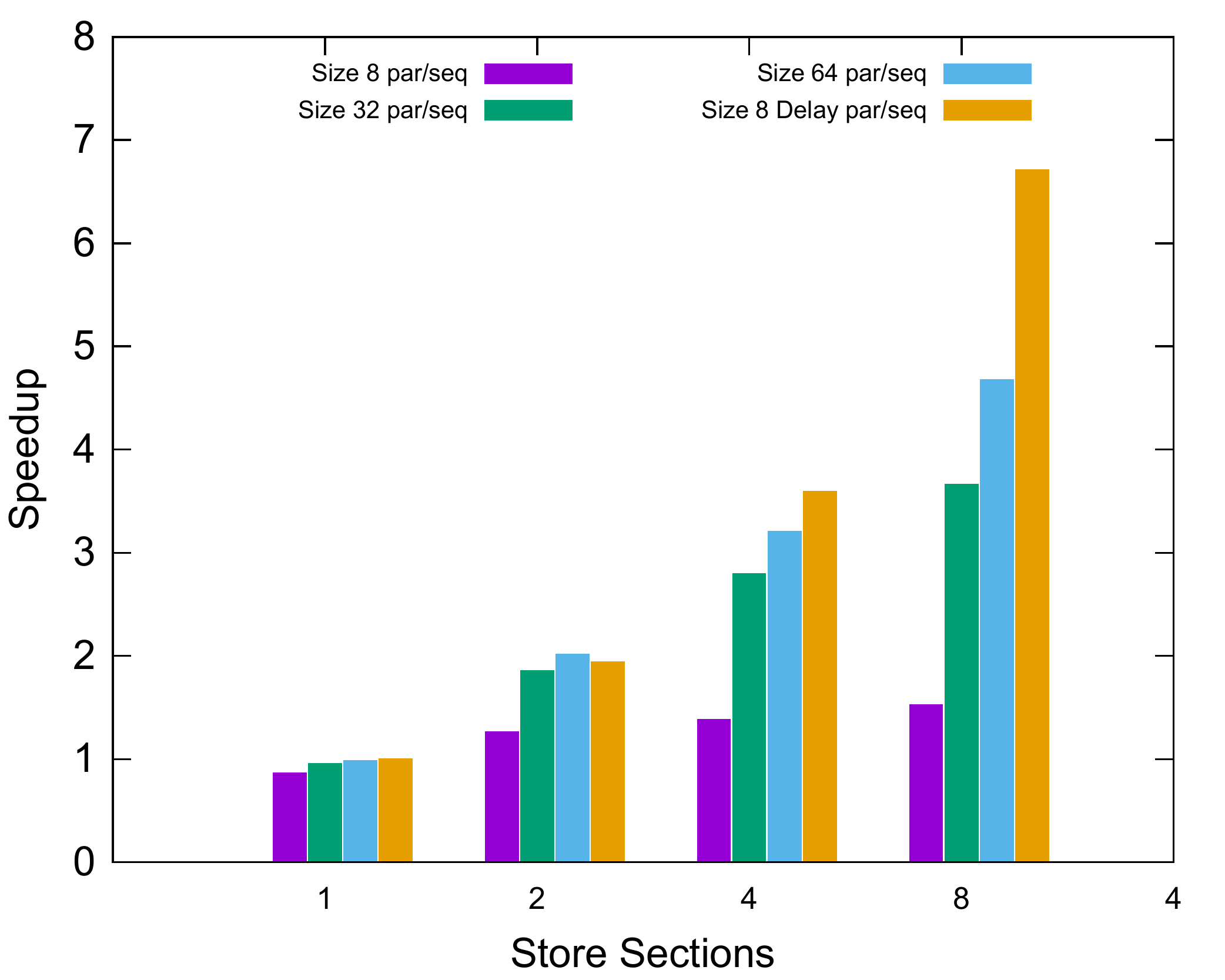}}
      \caption{Speedup with varying parallel resources and work.} \label{fig:throughput:speedup}
    \end{minipage} \hfill
    \begin{minipage}{0.32\linewidth}
    \centerline{\includegraphics[width=0.95\linewidth]{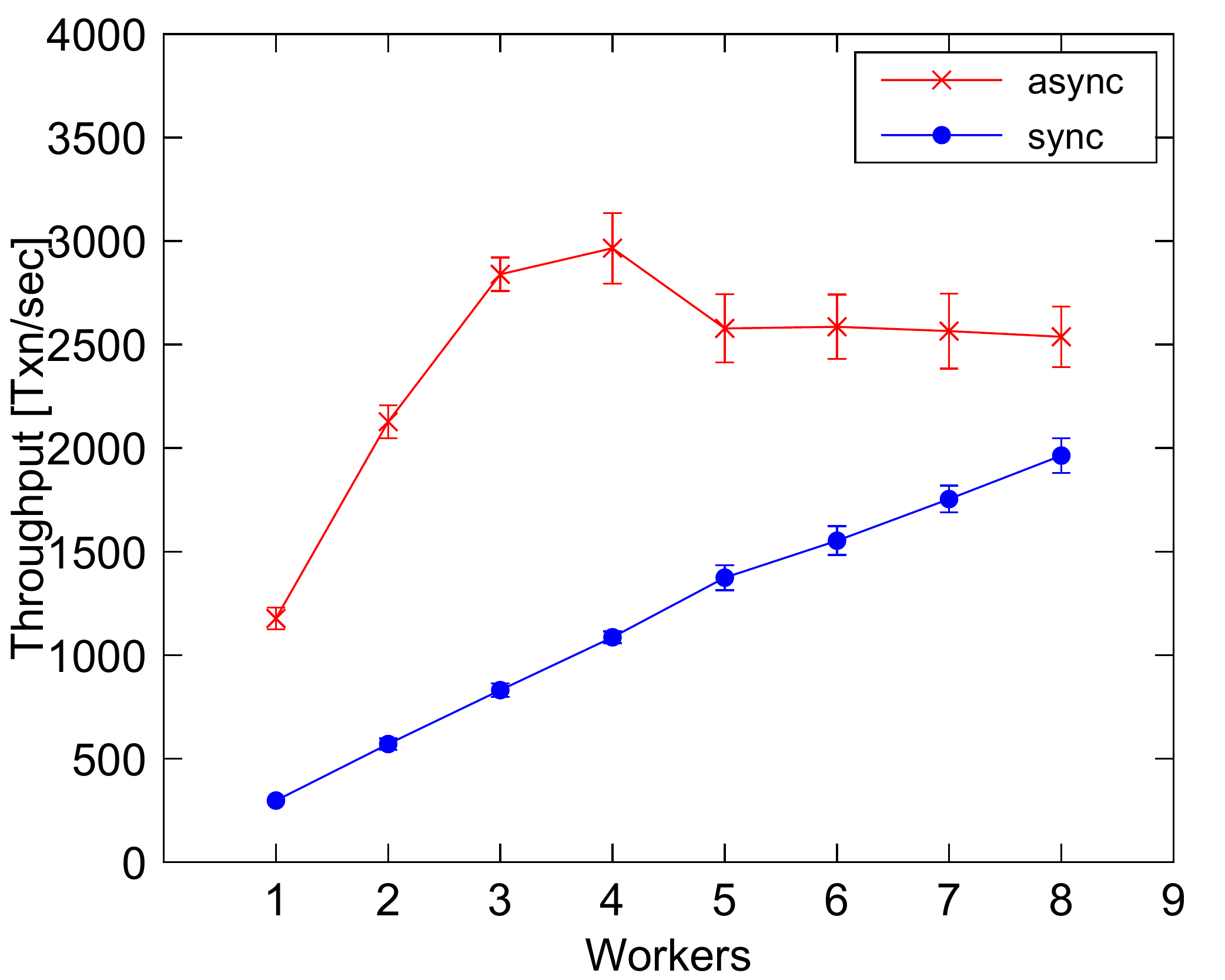}}
    \caption{Effect of load on throughput.} \label{fig:throughput:scalability}
    \end{minipage}
    \begin{minipage}{0.32\linewidth}
    \centerline{\includegraphics[width=0.95\linewidth]{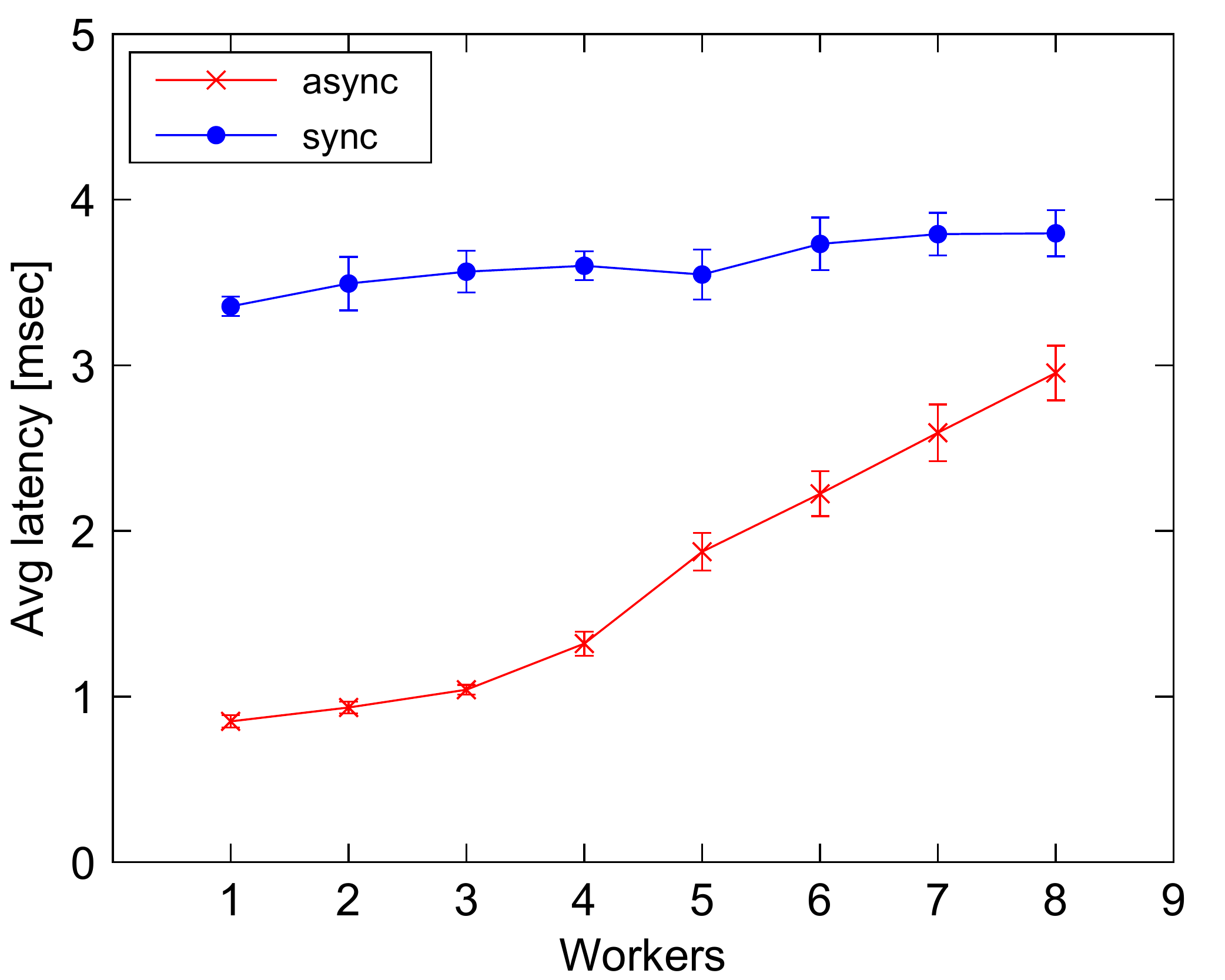}}
    \caption{Effect of load on latency.} \label{fig:latency:scalability}
    \end{minipage}
\end{figure*}

To characterize the limits of asynchronicity gains we can achieve with \reactdb, we evaluate the speedup obtained for the \textsf{async} deployment compared to the \textsf{sync} deployment of the SmartMart application as we vary the available parallel resources. We employ again a \emph{single worker} to aid in interpretability. We kept the total number of items in an order fixed and varied the number of store sections that those items are ordered from. Namely, when the total number of items in an order is $N$ and the total number of store sections that items are ordered from is $k$, then $N/k$ items are ordered from each store section. 

Figure~\ref{fig:throughput:speedup} shows the throughput speedups for our experiment with three different sizes of N, namely 8, 32, and 64, which are represented by \textsf{Size N par/seq} bars. As an extra control, we created another variant of variable discount computation where we replaced the scan over the history by an artificial delay of~3 msec created by random number generation to simulate complex prediction calculations. Such complex calculations would increase the ratio of parallel to sequential work without increasing the database footprint and the sequential commit cost. The variant is represented by the \textsf{Size 8 Delay par/seq} bar in Figure~\ref{fig:throughput:speedup}. We increased the delay by this specific amount so as to obtain close to 7x speedup, which we pre-calculated based on the ratio of sequential to parallel work in the interactions. Note that we computed the throughput speedup by calculating the ratio of \textsf{async} and \textsf{sync} throughput. 

We can see that at one store section, we have speedups of slightly less that one. This effect arises due to the lack of asynchronous execution and the small overhead of dispatch. As we increase the number of store sections, however, we can see that the speedups obtained increase for all transaction sizes. However, the increase is more pronounced in variants where the ratio of parallel to sequential work in the transactions is larger. Note that this effect is what is expected by application program structure, since a non-trivial fraction of the complex interaction logic, including \texttt{add\_items} and parts of \texttt{checkout}, is sequential. 

At the maximum available parallelism of eight store sections, the speedup of \textsf{Size 8 par/seq} is 1.52, the speedup of \textsf{Size 32 par/seq} is 3.66, the speedup of \textsf{Size 64 par/seq} is 4.67, and the speedup of \textsf{Size 8 Delay par/seq} is 6.8. According to Amdahl's law, in order to get a speedup of 7, 5, 4, 3 and 2 with 8 parallel resources, the parallelizable work must be 98\%, 91\%, 86\%, 76\% and 57\% of the entire work, respectively. To further drill down into potential sources of overhead, we also profiled the commit costs for \textsf{Size 8 seq/par} and \textsf{Size 8 seq/Delay par}, which were almost the same for both. These costs come to \textasciitilde68.2 $\mu$sec and \textasciitilde93.8 $\mu$sec for the \textsf{sync} and \textsf{async} versions, respectively. For comparison, the latency of the sequential \texttt{add\_items} function for \textsf{Size 8 seq/par} in \textsf{sync} and \textsf{async} was \textasciitilde125 $\mu$sec, corresponding to one-fourth the entire interaction latency.

This highlights the potential of asynchronicity gains that can be leveraged over parallel hardware as the amount of complex parallelizable patterns in transactional code increases and the overhead becomes a smaller fraction of total costs.

\subsection{Are asynchrony gains affected by inter-transaction parallelism in \textsf{SmartMart}?}
\label{sec:exp:scaleup}
\begin{figure*}[!t]
      \begin{minipage}{0.32\linewidth}
      \centerline{\includegraphics[width=0.95\linewidth]{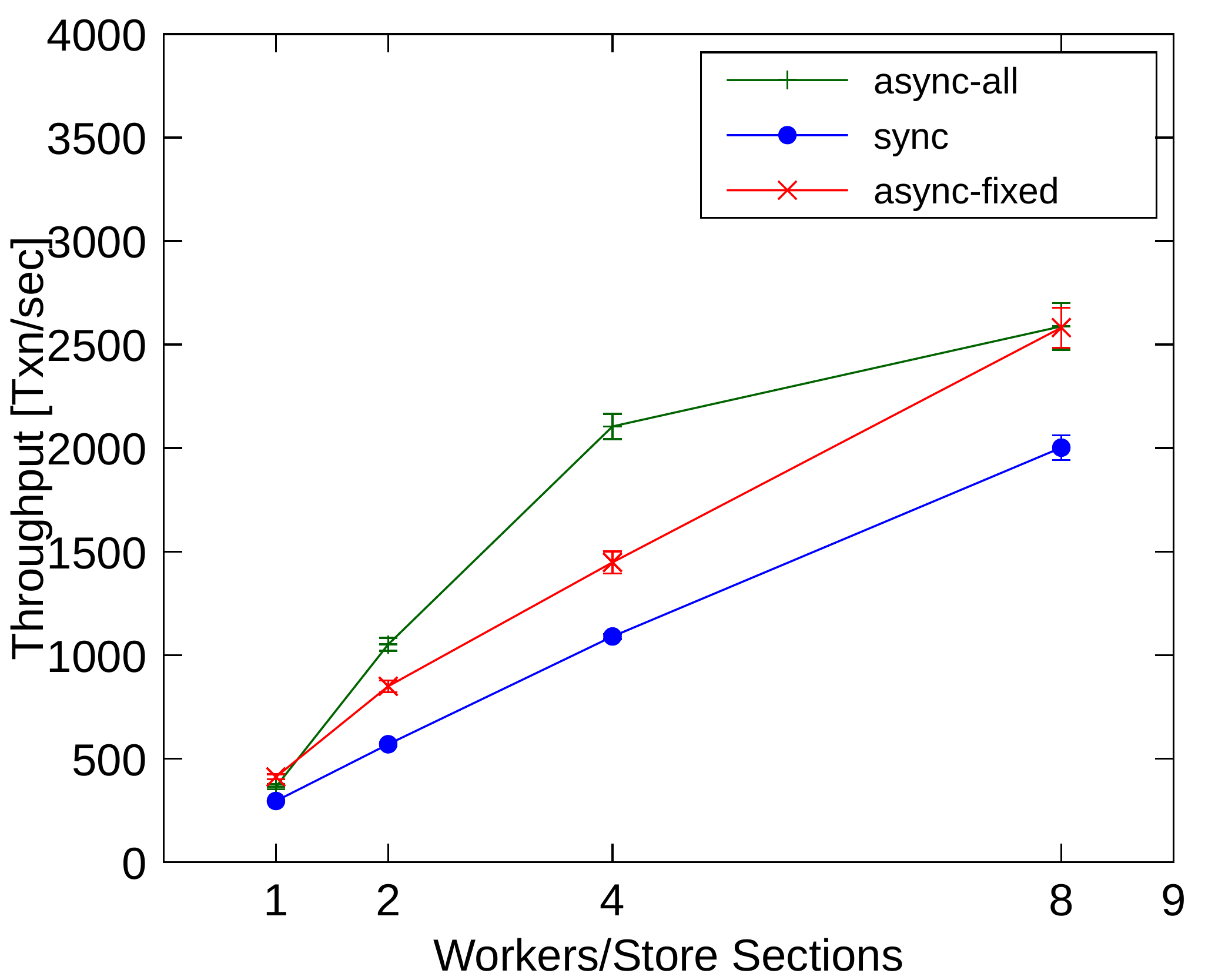}}
      \caption{Throughput scale-up on both workers and store sections.} \label{fig:scalability:throughput}
    \end{minipage} \hfill
    \begin{minipage}{0.32\linewidth}
      \centerline{\includegraphics[width=0.95\linewidth]{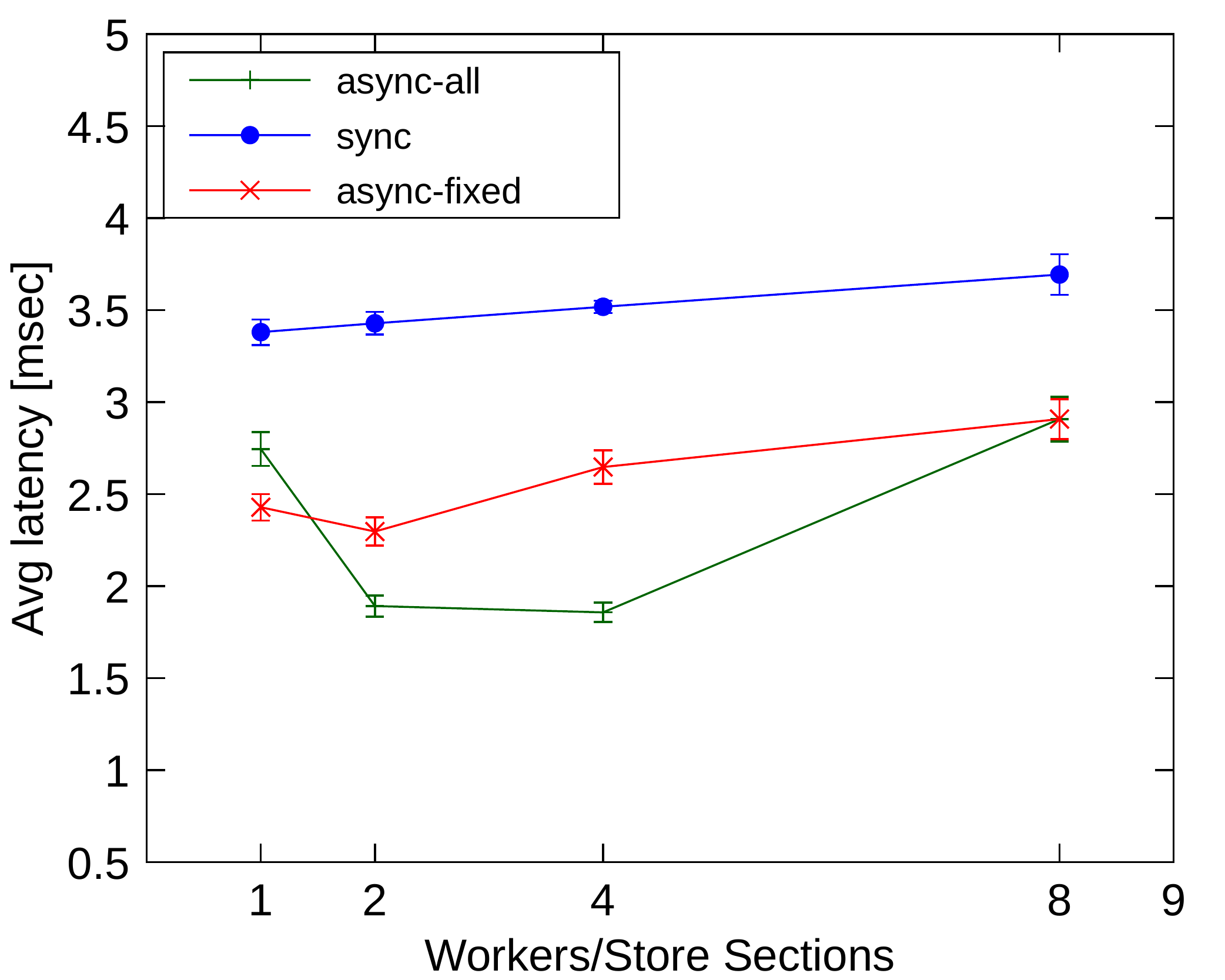}}
      \caption{Latency in scale-up of both workers and store sections.} \label{fig:scalability:latency}
    \end{minipage} \hfill
    \begin{minipage}{0.32\linewidth}
      \centerline{\includegraphics[width=0.95\linewidth]{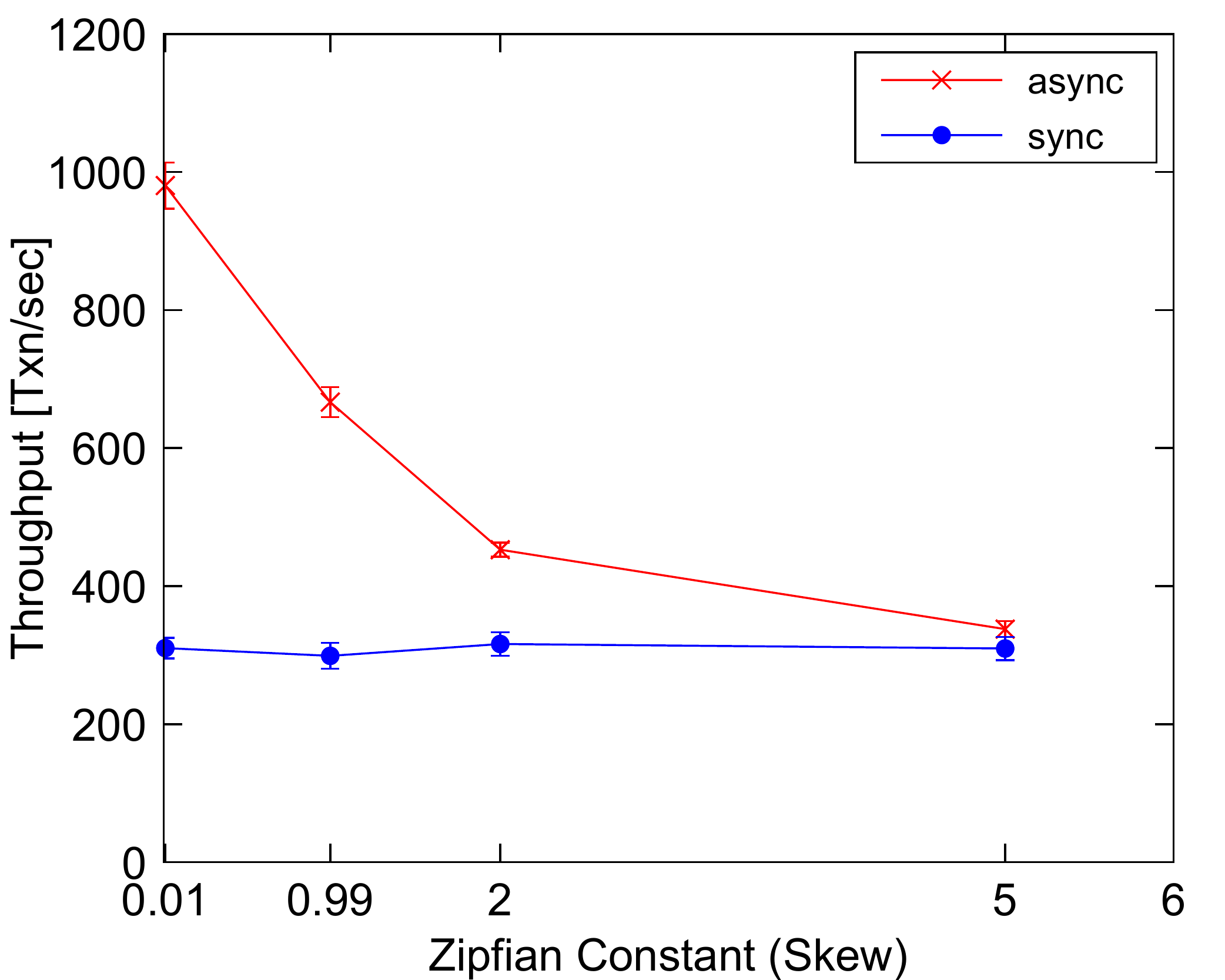}}
      \caption{Effect of skew on throughput.} \label{fig:skew:throughput}
    \end{minipage} \hfill
\end{figure*}

In this section, we study the interference effects due to inter-transaction parallelism (concurrent workers) on the gains from intra-transaction parallelism. By keeping parallel resources fixed (store sections) and increasing inter-transaction parallelism, we can study how load accentuates the interference effects on intra-transaction parallelism (see \emph{Increasing Load per Store Section}). On the other hand, by increasing both parallel resources and inter-transaction parallelism such that the load per parallel resource is constant, we can study how intra-transaction parallelism would be affected by interference effects upon scaleup~\cite{DeWittG92:Paralleldb} (see \emph{Constant Load per Store Section}). 

\minisec{Increasing Load per Store Section}
By gradually increasing the number of concurrent workers while keeping the number of store sections fixed, we study the effect of load on the benefits of asynchronicity observed previously. We keep the work fixed to an order size of 32, corresponding to an order across 8 store sections and 4 items from each store section. We increase the number of workers, carts and customers in the experiment. Figures~\ref{fig:throughput:scalability} and~\ref{fig:latency:scalability} show the throughput and latency observed. While \textsf{sync} exhibits excellent throughput scalability as we increase the number of workers, the throughput of \textsf{async} scales well until three workers and then degrades before roughly stabilizing. This is because at three workers, the \mbox{\texttt{Store\_Section}} reactors are close to full resource utilization (CPU cores at~88\%), maxing out at four workers and then becoming the bottleneck. The resulting effect of queuing can also be seen in the latency measurements, where the latency increases dramatically after four workers. 

Despite the queuing effects, \textsf{async} still outperforms \textsf{sync} because of the amount of physical resources being utilized by it, namely 16 cores with intra-transaction parallelism as opposed to 8 cores in sequential execution. We did not perform measurements for more than 8 workers, since the hardware does not have enough physical cores to sustain our setting for \textsf{async}. Nevertheless, we would expect a crossover with \textsf{sync} as load increases. In short, asynchronicity can bring both throughput and latency benefits over a traditional synchronous strategy when load in the database is light to normal and transactions exhibit parallelism. Our results are similar in nature to the earlier findings of the effect of load on asynchronous execution of OLTP transactions (TPC-C) in \reactdb~\cite{0001S18:ReactDB}.

During this experiment, we observed abort rates of $\sim$5-7\% despite the small amount of actual logical contention on items. This happens because the OCC protocol of Silo aborts transactions if the version numbers of nodes scanned change at validation time, caused in our experiments by tree splits due to inserts.

\minisec{Constant Load per Store Section}
In this experiment, we increase concurrent workers and physical resources at the same time, thus ensuring that the load on every store section is the same whether we employ one or more workers. This setup corresponds to the classic notion of scaleup~\cite{DeWittG92:Paralleldb}.

We keep the work fixed to an order size of 32 and vary the store sections that the order spans. For one worker, all the items are ordered from a single store section; for two workers, 16 items are ordered from each store section by each worker across two store sections; for eight workers, four items are ordered from each store section by each worker across eight store sections. Thus, we carefully control the load on each store section, with a total order size of 32 items irrespective of the number of workers. For clarity, we only chose those values of workers of which 32 is a perfect multiple in order to generate identical load across store sections.

Figures~\ref{fig:scalability:throughput} and~\ref{fig:scalability:latency} show the throughput and latency observed. We can see that \textsf{sync} exhibits excellent throughput and latency scalability. This is expected because the \textsf{sync} deployment is agnostic to intra-transaction parallelism and interference effects. We present two variants for \texttt{async}. In \textsf{async-fixed}, the store sections for the order are the same across all workers. For \textsf{async-all}, the store sections for the order are chosen uniformly at random from the eight store sections. Thus, \textsf{async-fixed} corresponds to the classical setup for scaleup, while \textsf{async-all} allows extra resources to be leveraged by intra-transaction parallelism when the load on the system is light to normal.

In Figure~\ref{fig:scalability:throughput}, \textsf{async-fixed} also demonstrates excellent scalability and reaches the same peak throughput and latency at eight workers as the \textsf{async} deployment in Figure~\ref{fig:throughput:scalability}. By contrast, \textsf{async-all} shows a much higher throughput gain until 4 workers before converging to the same value as that of \textsf{async-fixed} at eight workers. This effect arises because \textsf{async-all} benefits at first from lack of interference across transactions, since store sections are chosen uniformly at random. As the number of store sections in the order increase, the interference effects grow, and \textsf{async-all} converges at eight workers to the throughput of \textsf{async-fixed}. This effect can be clearly seen in the latency curve of \textsf{async-all} in Figure~\ref{fig:scalability:latency}, which initially shows benefits from intra-transaction parallelism before queuing delays due to increased load cause latency to degrade. 

We observe that the performance of \textsf{async-fixed} is higher than that of \textsf{sync}. This is expected because of the asynchronous invocations on store sections and the greater amount of parallel resources available, e.g., one core dedicated to carts and another core to a store section even for one worker. In addition, since a transaction spans the same physical cores, we conjecture that this effect reduces the amount of cross-core traffic. During this experiment, we observed peak abort rates of \textasciitilde5-7\% at eight workers, which is identical to the previous experiment with varying load per store section.

\subsection{What is the effect of skew in gains from asynchrony in SmartMart?}
\label{sec:exp:skew}
In this section, we study the effect of skew on the benefits of intra-transaction parallelism. For this experiment, we chose a single worker and kept the work fixed. We use a zipfian distribution to select the store sections that an item is chosen from and vary the zipfian constant to model skew on a store section. Figure~\ref{fig:skew:throughput} presents the throughput for an order size of 32 items across eight store sections. We also experimented with order sizes of 64 and 8 with delay (as used in Section~\ref{sec:exp:speedup}). Since we found the shape of the curves to be identical to the ones presented here, we omit these results for brevity.

We can see that as we increase the zipfian constant from 0.01 (no skew) to 0.99 (heavily skewed), the throughput of \textsf{async} decreases. This is because the amount of work across store sections is no longer balanced, thus introducing a dependency on the execution cost on the store section with the most orders (increased depth). The throughput of \textsf{sync} is agnostic to skew, since it does not utilize any intra-transaction parallelism. Despite a highly skewed workload when the zipfian constant is 0.99, \textsf{async} still outperforms \textsf{sync}. Even though some store sections are more popular than others, all the store sections are still utilized, leading to intra-transaction parallelism benefits. In other words, while there is a lack of balance across store sections, the result is still better than sequential execution. As we increase the zipfian constant to 5.0, however, the distribution becomes more and more skewed so that all the items in the order are selected from a single store section. At this point,  the performance of \textsf{async} converges to that of \textsf{sync}.

\subsection{Discussion}
In light of our evaluation, we make the following observations and reflections about the benefits of intra-transaction parallelism:\\
\begin{sloppypar}

\minisec{(1) OLTP with complex application logic can benefit from intra-transaction parallelism} We observe that OLTP applications with complex application logic intertwined with write-heavy data accesses can benefit from parallel execution on modern multi-core hardware. Gains from speedup were highest when the computational footprint of the application logic was higher than its data footprint, owing to commit overheads. However, even in the presence of data footprint in application logic generated by reading \textasciitilde100s of records, using intra-transaction parallelism was beneficial compared to sequential execution of individual transactions. This observation is significant given modern trends of deploying machine learning models in applications and their latency requirements for responsiveness~\cite{CrankshawWZFGS17:Clipper}.\\

\minisec{(2) \reactdb\ has low overheads for transactions with higher computational footprint than data footprint, but can be optimized further} For application logic with a high computational footprint, \reactdb's implementation mechanisms are efficient enough to provide speedups closer to theoretical limits estimated from Amdahl's law. Despite promising speedups for application logic with high data footprint, commit overheads diminish the gains from parallelism. We observe that this can potentially be addressed by parallelizing the linear 2PC protocol, using early lock release, and by lowering concurrency control overheads using techniques for read-mostly transactions~\cite{WangJFP17:SSN-ReadMostly}, to name a few possibilities.\\

\minisec{(3) Speedup in an OLTP workload with intra-transaction parallelism benefits not only latency, but also throughput} Parallel execution of an OLTP application with complex application logic over hardware resources with sufficient parallelism benefited both throughput and latency. The effects of higher performance result from better resource utilization than sequential execution of OLTP applications, but can only be observed as long as the hardware is not overloaded by interference effects of inter-transaction parallelism.\\

\minisec{(4) The containerization architecture of \reactdb\ allows for control of parallel and sequential execution of transactions, but only at deployment time}
The containerization mechanism of \reactdb\ allows assignment of reactors to containers such that programs accessing reactors in the same container are executed sequentially, despite being invoked asynchronously. This mechanic provides a flexible mechanism to enable and disable parallel execution of programs at deployment time. As such, the system can be configured to target workload characteristics based on expected load conditions and available hardware parallelism. However, the right configuration for a given scenario needs to be provided at database deployment time. It remains to be explored if adaptation of a deployment policy for best performance can be achieved dynamically depending on load conditions.\\

\minisec{(5) There is potential room to implement durability so that speedups would still be visible on modern hardware for applications with complex logic}
Since \reactdb\ currently does not support durability, we could not measure its impact on the gains of intra-transaction parallelism. We expect that a mechanism for durability will increase commit overhead, thus reducing potential intra-transaction parallelism gains. However, we have observed scenarios with complex application logic comprising large computational footprint or with large data footprint from reads but conventional OLTP data footprint for writes where raw performance leaves room for efficient durability implementations. We conjecture that logging costs could be made to range between a few 10-100 $\mu$sec depending on the modern hardware employed (viz. persistent memory and SSDs)~\cite{Dragojevic:2015:FaRM}. In such a case, intra-transaction parallelism speedups would still be visible in the scenarios observed where execution cost would then dominate logging overheads.   
\end{sloppypar}

\section{Related Work}
\label{sec:related}
Evaluation of transaction processing performance has a rich history, and a comprehensive review exceeds the scope of this paper. As a consequence, we narrow our focus in this section to contrast our study with the main categories of previous evaluation work of parallelism in transactions. 

Many early studies evaluated parallel transaction processing architectures and their impact on concurrency control schemes using performance modeling techniques~\cite{AgrawalCL87:CC-perf-model, CareyL88:CC-perf, CareyL89:CC-perf-more, BhideS88:CC-perf-arch}. Contemporary studies also went further than performance modeling and performed experimental evaluation with nascent parallel database systems~\cite{DeWittGS88:Gamma-perf, HuangSRT91:OCC-perf, ThakkarS90:Oltp-perf-smp, Group88:Debit-credit-perf}. In contrast to our work, all of these simulation and experimentation studies focused on evaluating the performance of disk-based database architectures and their concurrency control mechanisms under synthetic data-access oriented benchmarks, with the exception of~\cite{ThakkarS90:Oltp-perf-smp, Group88:Debit-credit-perf} that used the TP1 benchmark~\cite{Anonetal85:TP1}. Moreover, these early studies only considered intra-transaction parallelism achieved through query speedups by data partitioning~\cite{DeWittGS88:Gamma-perf}. By contrast, our work revisits the effects of the core factors affecting parallelism in the setting of actor-relational database systems with parallelization of arbitrary intra-transaction control flow in a modern OLTP application with complex application logic having task parallelism. 

Taking note of the popularity of multi-core machines, cloud computing infrastructure, and in-memory database systems, recent performance studies have focused on evaluating concurrency control architectures in a distributed infrastructure~\cite{HardingAPS17:CC-perf-distributed} and on single-node multi-core systems~\cite{YuBPDS14:CC-perf-multicore-simulator}. A separate line of recent work has also focused on evaluating micro-architectural effects of executing OLTP workloads on in-memory database systems~\cite{SirinTPA16:micro-arch-perf-oltp}. All of these studies focus on evaluating inter-transaction parallelism or high-volume transaction processing and its impacts on database architectural components. By contrast, our study focuses on the benefits of intra-transaction parallelism on multi-core machines. In addition to going beyond query-level data parallelism~\cite{LeisBK014:MorselDrivenParallelism, BarthelsAHSM17:DistributedJoin1000Cores, BalkesenATO13:SortvsHashRevisited, GicevaARH14:QueryPlanDeploymentMultiCores, HanKLLM08:Query-par}, to the best of our knowledge, our study is the first to perform such an evaluation in the context of actor-relational database systems.

\section{Conclusion}
This paper has studied the exploitation of intra-transaction parallelism by an actor-relational database system in an OLTP application with complex task-parallel application logic over multi-core hardware. A few distinctive features differentiate our evaluation. First, all intra-transaction parallelism is exposed to the database system by asynchronous programming constructs in an actor model. This parallelism is not limited to query-level, but extends to both data access and application logic within transactions. Second, we observe that variations in the classic factors affecting parallel efficiency, such as overhead, parallelizable work, interference, and skew, produce the expected effects on intra-transaction parallelization. These results indicate that the evaluated actor-relational database system, \reactdb, introduces manageable overhead in its implementation of transactional parallelism, suggesting that asynchronicity in actor-relational database systems can be effectively used to bring task-level parallelism gains to OLTP applications with complex application logic. Third, instead of pursuing an exclusive focus on maximum load and transaction throughput, our evaluation illustrates that in a range of light to normal load situations, asynchrony can lead to reduced transaction latencies by better use of parallel resources. These lower latencies can also engender higher throughputs than classic synchronous transaction execution strategies if idle parallel resources can be put to use by intra-transaction parallelization.       

An interesting area of future work is to study how asynchronicity can be exploited by an actor-relational database system in more complex multi-level parallelization scenarios, including a combination of multi-core and multi-node parallelism and potentially even many-core acceleration. Moreover, we believe that intra-transaction parallelism in complex application logic should be studied more generally in contexts encompassing but also going beyond actor-relational database systems.    

\bibliographystyle{ACM-Reference-Format}
\bibliography{actor-db-eval}
\end{document}